\documentclass[lettersize,journal]{IEEEtran}
\usepackage{amsmath,amsfonts}
\usepackage{algorithmic}
\usepackage{algorithm} 
\usepackage{array}
\usepackage{subfigure}
\usepackage{bm}
\usepackage[caption=false,font=footnotesize,labelfont=rm,textfont=rm]{subfig}
\usepackage{textcomp}
\usepackage{stfloats}
\usepackage{url}
\usepackage{verbatim}
\usepackage{graphicx}
\usepackage{cite}
\usepackage{amssymb}
\usepackage{booktabs}
\usepackage[table,xcdraw]{xcolor}
\usepackage{diagbox}
\usepackage{multirow} 
\usepackage{adjustbox}
\usepackage{enumitem}

\begin{document}
\title{Electromagnetic Neural Network for Direction-of-Arrival Estimation}
\author{Shining Lin, Jiancheng An,~\IEEEmembership{Senior Member,~IEEE,} Lu Gan,~\IEEEmembership{Member,~IEEE,} Victor C. M. Leung, \IEEEmembership{Life Fellow,~IEEE,}
Mehdi Bennis, \IEEEmembership{Fellow,~IEEE,} M\'erouane Debbah,~\IEEEmembership{Fellow,~IEEE,} and
Tie Jun Cui, \IEEEmembership{Fellow,~IEEE} 
\thanks{This work was supported by the National Natural Science Foundation of China (NSFC) under Grant 62471096. This article was presented in part at the IEEE International Conference on Acoustics, Speech and Signal Processing (ICASSP), Hyderabad, India, 2025 \cite{ICASSP_2025_Lin_UAV}. \emph{(Corresponding Author: Jiancheng An.)}}
\thanks{S. Lin and L. Gan are with the School of Information and Communication Engineering, University of Electronic Science and Technology of China (UESTC), Chengdu 611731, China. L. Gan is also with the Yibin Institute of UESTC, Yibin 644000, China (e-mail: 202221011710@std.uestc.edu.cn; ganlu@uestc.edu.cn).}
\thanks{J. An is with the School of Electronic Science and Engineering, University of Electronic Science and Technology of China (UESTC), Chengdu, 611731, China (e-mail: jiancheng.an@uestc.edu.cn).}
\thanks{V. C. M. Leung is with the College of Computer Science and Software Engineering, Shenzhen University, Shenzhen 518060, China, and also with the Department of Electrical and Computer Engineering, The University of British Columbia, Vancouver, BC V6T 1Z4, Canada (e-mail: vleung@ieee.org).}
\thanks{M. Bennis is with the Centre for Wireless Communications, University of Oulu, Oulu 90570, Finland (e-mail: mehdi.bennis@oulu.fi).}
\thanks{M. Debbah is with the Research Institute for Digital Future, Khalifa University, 127788 Abu Dhabi, UAE (e-mail: merouane.debbah@ku.ac.ae) and also with CentraleSupelec, University Paris-Saclay, 91192 Gif-sur-Yvette, France.}
\thanks{T. J. Cui is with the Institute of Electromagnetic Space, and the State Key Laboratory of Millimeter Wave, Southeast University, Nanjing 211189, China (e-mail: tjcui@seu.edu.cn).}\vspace{-1cm}}
\markboth{DRAFT}{DRAFT}
\maketitle
\begin{abstract}
Accurate and real-time direction of arrival (DOA) estimation is crucial for beamforming in unmanned aerial vehicle (UAV) communication systems. However, the existing high-precision DOA estimation algorithms encounter high computational complexity when implemented on a UAV with on-board signal processing constraints. To tackle this issue, an electromagnetic neural network (EMNN) is developed for DOA estimation, which is capable of generating the angular spectrum of the incident signal based solely on amplitude observation. Specifically, the proposed EMNN consists of two components: a stacked intelligent metasurfaces (SIM) is mounted on the UAV, and each meta-atom is an artificial neuron that can process signals in the electromagnetic domain with low energy consumption and ultra-fast computing speed. Furthermore, a fully connected layer is cascaded to process the received amplitude signal, enhancing the non-linear extraction and representational ability of EMNN. Moreover, to reduce the computational complexity and observation snapshots required for high-resolution DOA estimation, we develop a hierarchical DOA estimation framework, which involves two stages for conducting coarse and fine DOA estimation, respectively. For each stage, EMNN is trained on randomly generated training samples and their corresponding spectra to achieve the desired estimation goal. Finally, the simulation results validate that the proposed EMNN achieves approximately 13 dB gain in classification error reduction over the conventional beamforming (CBF) method in dual-signal scenarios, albeit its lower cost and radio frequency (RF)-related power consumption.
\end{abstract}


\section{Introduction}
\IEEEPARstart{T}{he} realization of seamless connectivity is widely acknowledged as a key feature of future sixth-generation (6G) networks, with unmanned aerial vehicle (UAV) technology being a potent means to achieve this goal. In Release 18, 3GPP has introduced non-terrestrial networks as an effective complement to terrestrial communication networks\cite{3GPPTR}. Capitalizing on their advantages of broad and adaptable deployment, UAV networks offer: i) Extensive source localization and surveillance capabilities\cite{cong2023crb}; ii) Emergency communication capabilities for disaster relief scenarios\cite{JSAC_2021_Do-Duy}; iii) Routing and broadcasting services to enhance quality of service (QoS) for terrestrial networks\cite{mei2020cooperative}. In UAV communication systems, fast and accurate direction of arrival (DOA) estimation is a crucial task due to two reasons. First, it facilitates effective beam steering to enhance beamforming gain and increase the achievable communication rate\cite{miao2021location}. Second, the DOA provides a geometric property that can be leveraged for other sensing functionalities such as localization\cite{lin2020multiple,cong2023crb}.

Common DOA estimation algorithms are based on observations from multiple sensors. Initially, receiving beamforming algorithms were employed for DOA estimation\cite{krim1996two}, and under specific array arrangements, they could be computed rapidly using the fast Fourier transform (FFT) algorithm. However, the resolution of beamforming algorithms is limited by the Rayleigh criterion. To overcome this limitation, several high-resolution algorithms have been proposed. Among them, the multiple signal classification (MUSIC)\cite{schmidt1986multiple} and the estimation of signal parameters via rotational invariance techniques (ESPRIT)\cite{roy1989esprit}, both based on the signal subspace, are the most representative. The former produces an angular spectrum by leveraging the orthogonality of the signal subspace and noise subspace. The latter utilizes the rotational invariance of the signal subspace. Moreover, on this basis, some variants have been introduced, such as Root-MUSIC\cite{barabell1983improving} and unified ESPRIT\cite{haardt1995unitary}, to reduce the complexity for searching spectral peaks and catering to coherent signal scenarios. Although traditional model-based methods can excel in certain scenarios, they come at a high computational cost, particularly when involving eigenvalue decomposition. Moreover, their performance drops substantially when there are limited data snapshots or in the face of hardware imperfections.

In recent years, there has been a growing trend in the development of DOA estimation algorithms based on advanced machine learning (ML) techniques\cite{papageorgiou2021deep,liu2018direction,elbir2020deepmusic,merkofer2022deep,AOM_2022_Chen_Artificial,NC_2025_Chen_Integrated}. Compared to traditional approaches, these data-driven ML methods exhibit greater adaptability to a broader range of scenarios and demonstrate enhanced robustness under challenging conditions such as the limited number of snapshots and low signal-to-noise ratio (SNR). There are primarily two types of ML-based DOA estimation algorithms. The first type directly learns the mapping between observed signals and their DOA, without explicit modeling. For instance, in\cite{papageorgiou2021deep}, DOA estimation was modeled as a multi-label classification task, which was addressed by employing a convolutional neural network (CNN). They demonstrated that the proposed CNN outperforms conventional methods in low SNR scenarios. In\cite{liu2018direction}, a multi-layer perceptron (MLP) was utilized to construct the angular spectrum, providing robustness against array imperfections. The second type integrates machine learning with traditional model-based methods. For instance, in\cite{elbir2020deepmusic}, a CNN network was utilized to learn the MUSIC spectrum from the covariance matrix.

\begin{table*}[t]
\centering
\caption{Comparison of various neural networks for DOA estimation}
\renewcommand{\arraystretch}{1.25}
\resizebox{1\textwidth}{!}{
\begin{tabular}{@{}ccccccc@{}}
\toprule
Network type &
 \begin{tabular}[c]{@{}c@{}}RF-related\\ hardware complexity\end{tabular} &
 Computation load &
 Operation latency &
 \begin{tabular}[c]{@{}c@{}}Number of observation\\ snapshots required\end{tabular} &
 \begin{tabular}[c]{@{}c@{}}Feature extraction\\ capability\end{tabular} &
 References \\ \midrule
ANN & High & High & High & Low & High & \cite{papageorgiou2021deep,liu2018direction,elbir2020deepmusic,merkofer2022deep,AOM_2022_Chen_Artificial,NC_2025_Chen_Integrated} \\
SIM & Low & Low & Low & High & Moderate & \cite{JSAC_2024_An_Two,gao2024super} \\
EMNN & Low & Low & Low & Moderate & High & $\star$\\ \bottomrule
\end{tabular}}\vspace{-0.5cm}
\end{table*}

Despite the advantages offered by ML methods, their speed and energy consumption remain constrained by digital computing units. Moreover, in UAV communication scenarios, the stringent constraints on UAVs' embedded hardware render it difficult to perform large-scale digital computations in real-time. Recently, stacked intelligent metasurfaces (SIM)\cite{lin2018all,liu2022programmable, arXiv_2025_Liu_Stacked}, as a novel computing device, have garnered attention due to their substantial advantages in terms of computational energy efficiency and latency when compared to traditional digital computing units. SIM consists of several customized diffraction layers, each composed of numerous meta-atoms that are capable of tuning incident electromagnetic (EM) waves. As EM waves propagate through these meta-atoms, they act as secondary sources, generating secondary waves that propagate to all meta-atoms in the subsequent layer based on the Rayleigh-Sommerfeld diffraction principle \cite{lin2018all}. The received signal at each meta-atom in the next layer is the weighted sum of signals radiated by all meta-atoms in the preceding layer. Viewing each meta-atom as a transmissive neuron, a SIM possesses an electromagnetic neural network (EMNN) architecture with controllable wave propagation for performing computing automatically. Early SIMs were fabricated from passive materials, such as 3D printed materials\cite{lin2018all} and polarization conversion units\cite{huang2023diffraction}. Their structures remain fixed once manufactured, constraining their adaptability for diverse tasks or dynamic environments. In recent years, reconfigurable metasurface technology has undergone rapid advancements \cite{arXiv_2024_An_Emerging, CL_2023_An_Tutorial,TWC_2025_An_Flexible, WC_2024_An_Codebook}, including reconfigurable transmission metasurfaces capable of dynamically manipulating the phase or amplitude of EM waves in real-time. Therefore, reconfigurable transmission metasurfaces have been developed as diffraction layers for SIM \cite{liu2022programmable,TAP_2024_Jia_High}. 
These metasurfaces are designed using meta-atoms that incorporate programmable components like power amplifiers \cite{liu2022programmable} and diodes \cite{TAP_2024_Jia_High}, significantly enhancing the adaptability of SIM.

Over the past years, the EM tuning capability of SIM has found diverse applications. For image recognition, in\cite{liu2022programmable}, grayscale images of MNIST handwritten digits were modulated into the transmission coefficient pattern of the input metasurface, and the designed 5-layer SIM achieved 90.76\% recognition accuracy. In wireless communications, leveraging the capability of SIM to effectively control radiation patterns \cite{OJCOMS_2025_Hassan_Efficient}, \textit{An et al.} \cite{TWC_2025_an_Stacked,ICC_2025_Liu_DRL} integrated a SIM with the radome of the base station, achieving multiuser downlink beamforming entirely in the wave domain. Later, this architecture was extended to the MIMO scenario by placing another SIM at the receiver, enabling the capability of MIMO precoding and combining in the wave domain\cite{AnSIM, an2024stacked, papazafeiropoulos2024achievable}. Furthermore, the SIM-based MIMO transceiver has garnered significant research attention across a diverse range of applications, including integrated sensing and communications \cite{WCL_2025_Niu_Stacked}, satellite communications \cite{WCL_Lin_2024_Stacked}, cell-free networks \cite{TWC_2025_Shi_Uplink}, secure communications \cite{TIFS_2025_Niu_Efficient}, and semantic communications \cite{WCL_2025_Huang_Stacked}, among others. For DOA estimation, in\cite{gao2024super}, a SIM was trained for focusing incident EM waves of different frequencies and DOAs onto corresponding positions on the detection plane. By doing so, DOA estimation can be executed by detecting the received peak power. Nevertheless, the resolution of this method is constrained by the number of sensors, and the constrained inference capability of the SIM further limits the precision of DOA estimation. Furthermore, in\cite{JSAC_2024_An_Two}, a SIM was used to generate the two-dimensional (2D) Fourier angular spectrum. By subtly varying the SIM's phase shifts to assemble the entire angular spectrum, the accuracy of DOA estimation was improved at the cost of increasing the number of observation snapshots, given the limitations in the number of sensors.

However, even though increasing the number of diffraction layers enhances the computing performance\cite{kulce2021all}, the intrinsic linearity of the input-output relationship in the overall architecture of the SIM constraints its inference capability, making it difficult to complete complex tasks that an artificial neural network (ANN) can handle independently. Therefore, to further enhance its versatility, nonlinear effects should be integrated into the EMNN architecture, making it closer to an ANN. To this end, significant efforts have been made by cascading metasurface layers with nonlinear amplitude measurement for achieving nonlinear activation functions in the analog domain. Then, a small number of digital hidden layers are involved for further processing the amplitude signals\footnote{To make a distinction, the traditional cascade of multiple metasurfaces is termed as SIM, and the neural network that introduces the nonlinear response and possesses wave-domain computing power is termed as EMNN. In certain cases, the EMNN considered in this paper is also termed as the hybrid optical-electric neural network (HOENN) \cite{chang2018hybrid,fan2023optical,qu2022all,zhou2021large}.}\cite{chang2018hybrid,fan2023optical,qu2022all,zhou2021large,mengu2020analysis,chen2023all,Science_2024_Gu_Direct}. This structure strikes flexible tradeoffs between computational load, hardware complexity, and inference capability. For instance, in\cite{mengu2020analysis}, \textit{Mengu et al.} explored the enhancement effects by cascading SIM with diverse digital neural networks, and in\cite{chen2023all}, SIM is cascaded with an analog electronic neural network comprised of photodiodes, resulting in a high-speed and energy-efficient EMNN.

\begin{figure}[!t]
\centering
\subfigure[Traditional sensing receiver]{
		\includegraphics[width=0.4\textwidth]{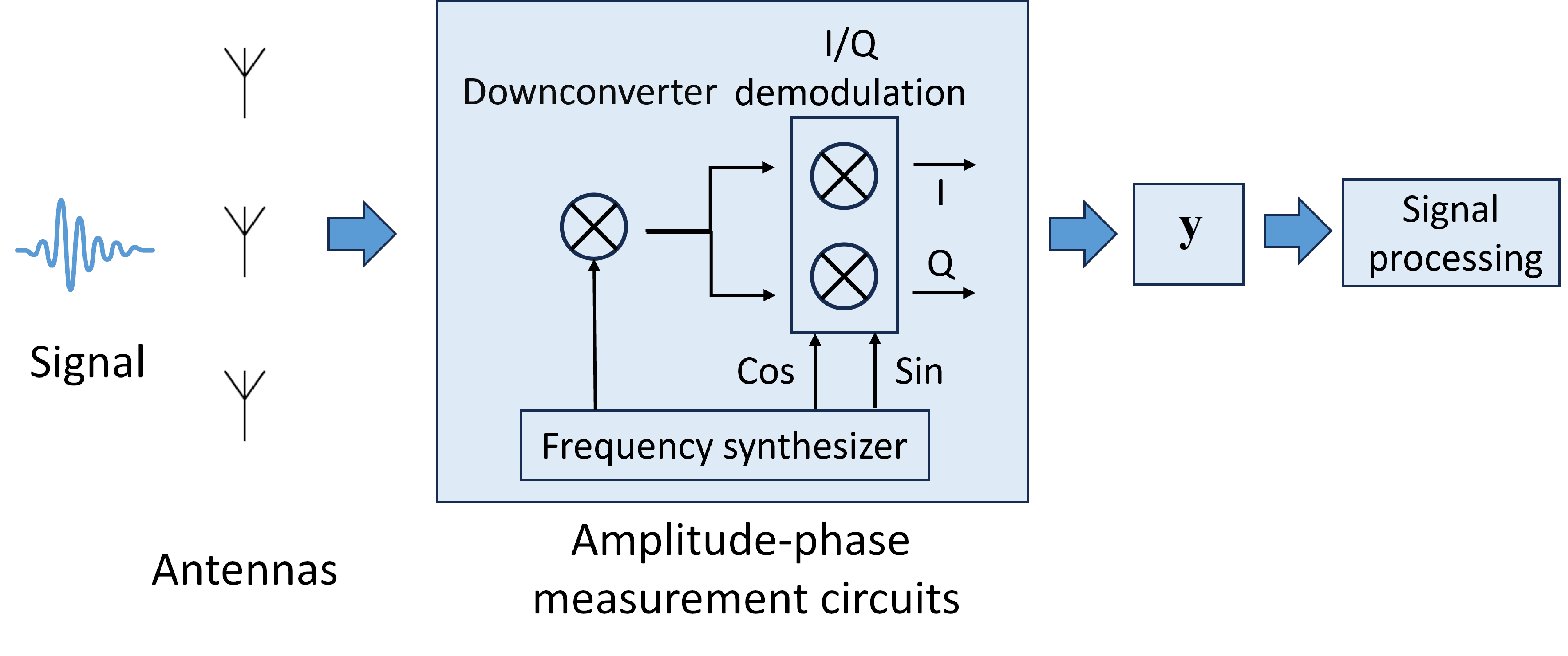}}
\subfigure[SIM-based sensing receiver]{
		\includegraphics[width=0.4\textwidth]{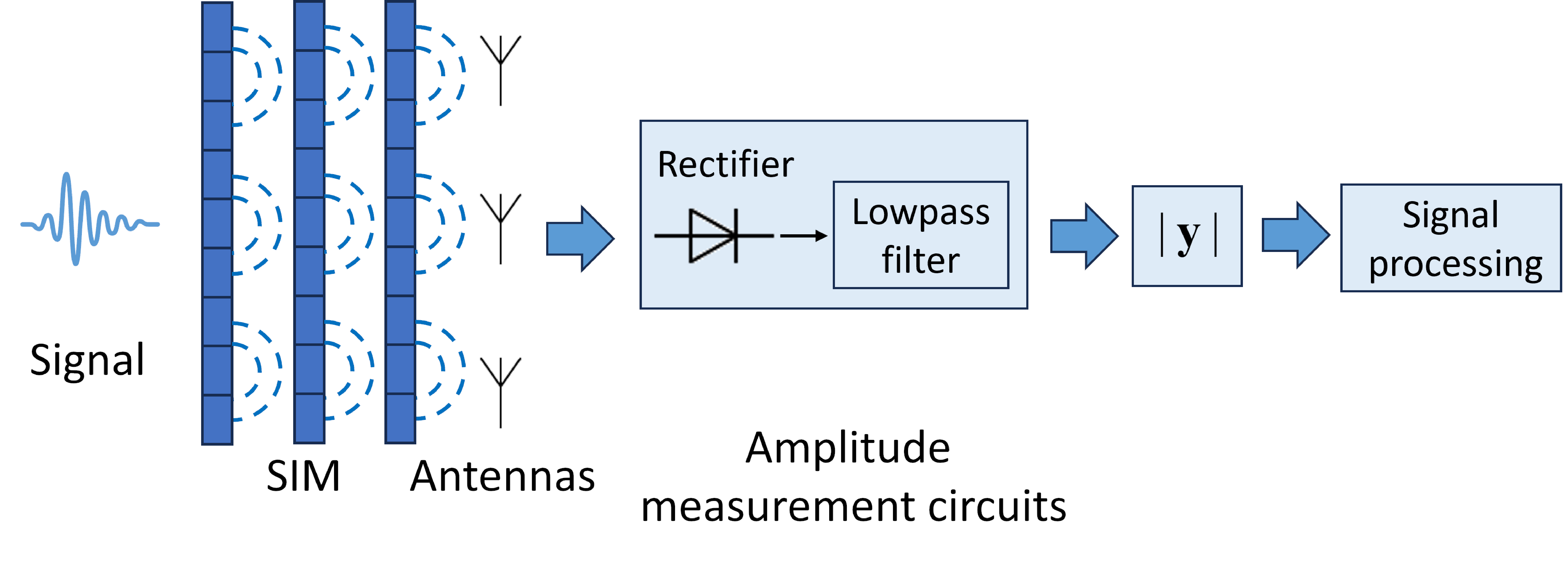}}
\caption{Comparison between traditional and SIM-based sensing receivers.}
\label{fig_1}\vspace{-0.5cm}
\end{figure}

Motivated by these observations, we develop a novel DOA estimation method using an EMNN, which comprises a SIM and a single fully connected layer. Since this represents the first attempt to utilize an EMNN for DOA estimation, the characteristics of various neural networks -- including ANN, SIM, and EMNN -- such as their computational load and operation latency during inference, are summarized in \textbf{Table I}. It is revealed that an EMNN, which consists of SIM-based analog hidden layers and digital hidden layers, enhances the computing capability of pure SIM. Compared to a digital ANN, the EMNN is capable of reducing RF-related power consumption and hardware complexity. Furthermore, the preprocessing by SIM further reduces the digital computational burden. Moreover, the EMNN requires fewer observation snapshots compared to SIM for implementing the DOA estimation task. Therefore, compared to other network types, the EMNN-based DOA estimation method is more suitable for embedded scenarios where latency, power consumption, and complexity are critical.

As shown in Fig. \ref{fig_1}(a), traditional DOA estimation methods based on amplitude-phase measurement require a frequency synthesizer for down-conversion and I/Q separation. During this process, I/Q balance and phase synchronization should be ensured across multiple antennas. In contrast, SIM-based sensing receivers only require relatively simple amplitude measurement circuits, such as the envelope detectors shown in Fig. \ref{fig_1}(b). We note that another class of sensing architectures employs EM lenses, which exploit their spatial focusing capability to map incident waves from different directions onto a focal arc for DOA estimation \cite{ICCECT_2023_Xu_Low}. However, a SIM can achieve stronger feature extraction capabilities through appropriate training, thereby realizing higher DOA estimation accuracy, while also offering greater placement flexibility (e.g., the receiving antennas are not restricted to the focal arc). Moreover, the programmable SIM-based sensing receiver can better adapt to diverse scenarios through various preset configurations, thereby achieving superior performance. Furthermore, for ease of comparison, the characteristics of these three sensing architectures are summarized in \textbf{Table II}.

\begin{table}[!t]
\centering
\caption{Comparison of various sensing receivers}
\renewcommand{\arraystretch}{1.25}
{\begin{tabular}{l||c|c|c}
\hline
 \diagbox{Metrics}{Receiver type} & 
 Traditional & 
 \begin{tabular}{@{}c@{}}EM lens-\\ based\end{tabular} & 
 \begin{tabular}{@{}c@{}}SIM-\\ based\end{tabular} \\
\hline
 \begin{tabular}{@{}l@{}}Information preservation\end{tabular} & Strong & Poor & Moderate \\
\hline
 Deployment flexibility & Flexible & Restricted & Flexible \\
\hline
 Cost & High & Low & Moderate \\
\hline
 Adaptability & Adaptable & Fixed & Adaptable \\
\hline
\end{tabular}}\vspace{-0.5cm}
\end{table}

More specifically, the major contributions of this work are outlined as follows: 
\begin{enumerate}[]
\item We design an EMNN for DOA estimation. Specifically, the DOA estimation process comprises three primary steps. Firstly, the SIM extracts the spatial distribution characteristics of incident EM waves and converts them into received power information at the receiving antenna array. Notably, all computations occur naturally as the EM signals propagate through the SIM at the speed of light, resulting in remarkably low energy consumption and latency. Subsequently, the power received at the antenna array is measured and connected to a fully connected layer to generate the angular spectrum. Finally, the DOA estimate is determined through peak detection. Compared to existing DOA estimation methods based only on SIM, EMNN requires fewer snapshots and, thus can be utilized to accommodate high-speed UAV scenarios.
\item A novel DOA estimation protocol based on amplitude measurement is developed.
This method is in sharp contrast to conventional DOA estimation approaches relying on amplitude-phase measurements. Specifically, it replaces the intricate and power-intensive RF chains with straightforward and energy-efficient power measurement circuits. Due to the preprocessing of the SIM, it can achieve satisfactory DOA estimation performance even when relying on the low-complexity energy detector, which makes it very suitable for UAV communications with embedded hardware constraints. Moreover, it solely requires the estimation of received amplitude rather than the covariance matrix, resulting in reduced computational complexity for smoothing noise over multiple snapshots.
\item To enhance the DOA estimation performance given the limited number of antennas and SIM scale, time-division multiplexing is utilized to expand the scale of EMNN. Specifically, by inputting the power received by antennas in different time blocks with different SIM phase shift configurations into the fully connected layer, the controllable parameters are proportionally increased without increasing hardware costs, while enhancing the inference capability of the EMNN.
\item To mitigate the demand for excessive observation snapshots and computational resources, a hierarchical DOA estimation framework, consisting of two steps, coarse estimation and fine estimation, is developed. In the coarse estimation phase, EMNN is employed to generate a low-resolution angular spectrum to identify the possible DOA range. Then, in the fine estimation phase, EMNN’s parameters are adjusted to generate a high-resolution angular spectrum within the reduced range, thereby facilitating the DOA estimation.
\item The Cramér-Rao bound (CRB) for a single source is derived for the SIM-based sensing receiver under amplitude measurement with single-snapshot constant-modulus signals. Simulation results are provided to investigate the impact of SIM training on the CRB.
\item Simulation results validate the feasibility of the EMNN for DOA estimation and the effectiveness of the proposed hierarchical estimation design. Specifically, the proposed hierarchical estimation protocol achieves a DOA estimation error of approximately 0.01$^{\circ}$. Furthermore, under certain circumstances, the proposed DOA estimation technique outperforms the conventional beamforming (CBF) method, showing considerable benefits in common multi-signal scenarios.
\end{enumerate}

\begin{table}[!t]
\caption{A list of major symbols and their definitions}
\centering
\renewcommand{\arraystretch}{1.25}
\begin{tabular}{c||l}
\hline
Symbol & Definition \\ \hline
$L$ & Number of metasurfaces \\ \hline
$N$ & Number of meta-atoms on each layer\\ \hline
$M$ & Number of antennas \\ \hline
$K$ & Number of incident signals \\ \hline
$G$ & Number of divided regions for coarse estimation \\ \hline
$V$ & Number of divided regions for fine estimation \\ \hline
$S$ & Number of snapshots for each SIM configuration \\ \hline
$Q$ & Number of SIM configurations used by EMNN \\ \hline
\end{tabular}\vspace{-0.5cm}
\end{table}

\textit{Notations}: Scalars, vectors, matrices, and sets are denoted by italic letters, lowercase bold letters, uppercase bold letters, and uppercase calligraphic letters, respectively; $\mathbb{C}^{M \times N}$ and $\mathbb{R}^{M \times N}$ represent the spaces of complex matrices and real matrices with dimensions $M \times N$; For a vector $\bf x$, $[{\bf x}]_i$, $||{\bf x}||_F$, $\text{diag}({\bf x})$ represent its $i$-th element, Frobenius norm, diagonal matrix with its elements as diagonal elements, respectively; ${\bf x}\in\mathcal{CN}({\bf m},{\bf R})$ indicates that the vector $\bf x$ follows a complex Gaussian distribution with mean ${\bf m}$ and covariance matrix $\bf R$; For a real vector ${\bf r} \in \mathbb{R}^{P\times1}$, $\min\{{\bf r}\}$ and $\max\{{\bf r}\}$ represent its minimum value and maximum value, respectively; $\bf z =\text{softmax({\bf r})}$ represents softmax operation, which is defined by $[{\bf z}]_p={e^{[{\bf r}]_p}}/{\sum_{p=1}^{P}e^{[{\bf r}]_p}}, p=1,\cdots,P$; For a matrix ${\bf A}$, $[{\bf A}]_{i,j}$, ${\bf A}^T$, and ${\bf A}^H$ denote its $(i,j)$-th element, transpose and conjugate transpose, respectively; $|\mathcal{A}|$ denotes the cardinality of set $\mathcal{A}$; ${\bf A}\otimes{\bf B}$ represents the Kronecker product of ${\bf A}$ and ${\bf B}$; $\mathcal{A}\cup \mathcal{B}$ and $\mathcal{A}\cap \mathcal{B}$ represent the union and intersection of sets $\mathcal{A}$ and $\mathcal{B}$, respectively; $|\cdot|$, $\mathbb{E}[\cdot]$, $\cos^{-1}(\cdot)$, real$(\cdot)$ represent the modulus operation, the expectation operator, the arccosine function, and the operation of taking the real part, respectively; $\frac{\partial y}{\partial x}$ represents the partial derivative of $y$ with respect to $x$; $j$ is the imaginary unit; $\bf 1$ represents an all-one matrix with appropriate dimensions. The major symbols used in this paper and their meanings are summarized in \textbf{Table III}.

\begin{figure}[!t]
\centering
\includegraphics[width=0.4\textwidth]{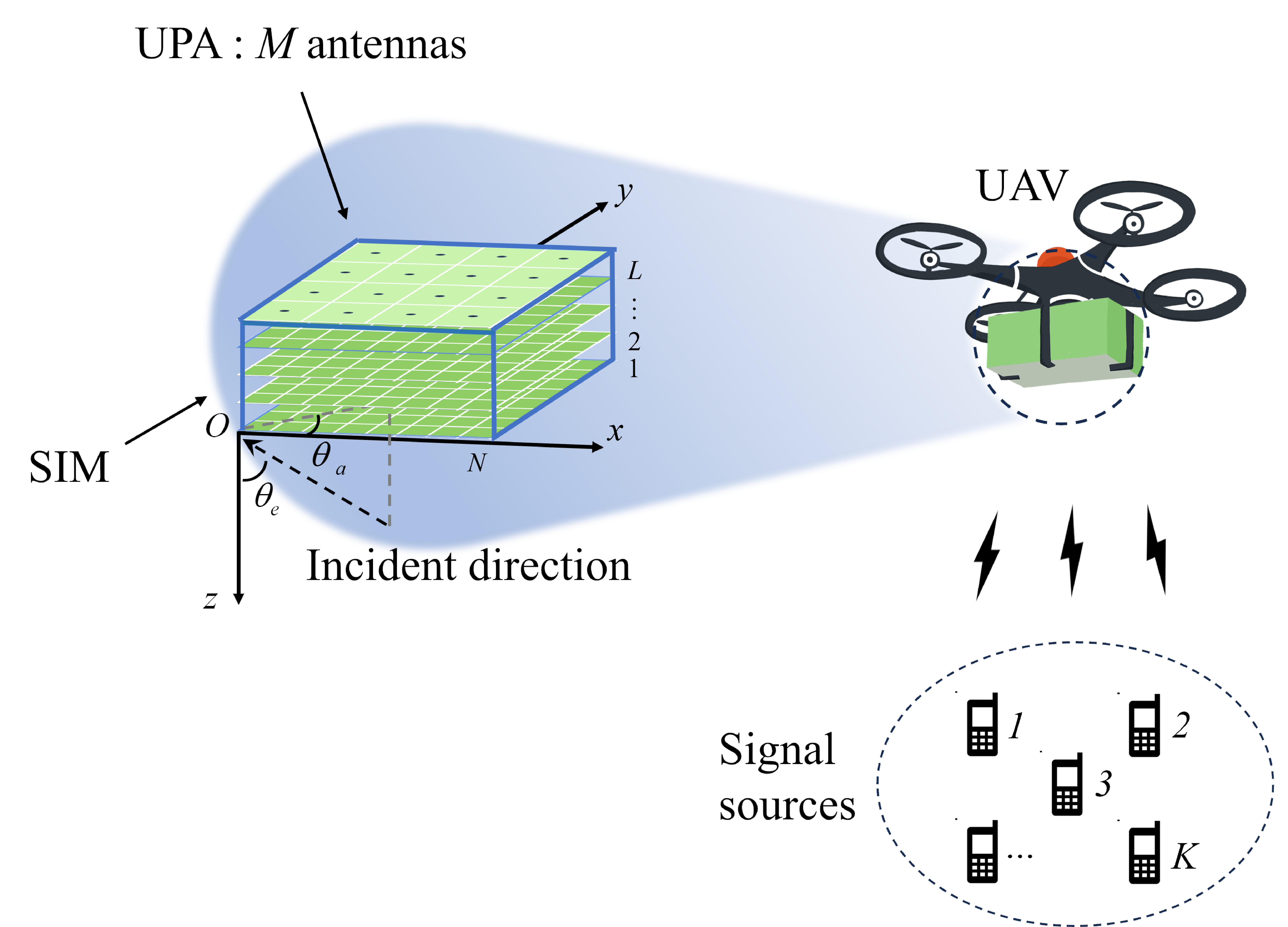}
\caption{A schematic of the considered system model, where a UAV is equipped with a SIM.}
\label{fig_2}\vspace{-0.5cm}
\end{figure}

\section{System Model}
As shown in Fig. \ref{fig_2}, we consider a UAV equipped with a uniform planar array (UPA) and a SIM for DOA estimation\footnote{Notably, the proposed EMNN-based method is not restricted to UAV scenarios; rather, it is well-suited for embedded platforms with constrained power consumption and hardware complexity.}. The SIM consists of many meta-atoms, whose EM responses can be reconfigured in real-time based on an attached smart controller. As a result, the UAV only performs simple computations based on the received signal power from the antenna array to obtain the angular spectrum. Specifically, the UPA consists of $M=M_x\times M_y$ antennas, where $M_x$ and $M_y$ represent the numbers of antennas in the x-axis and y-axis directions, respectively. The antenna spacing in the x-axis direction and y-axis direction is represented as $d_{x}^{\text{A}}$ and $d_{y}^{\text{A}}$, respectively. The SIM comprises $L$ layers of equidistant metasurfaces parallel to the UPA, each composed of $N=N_x\times N_y$ meta-atoms arranged in a uniform rectangular array, where $N_x$ and $N_y$ denote the number of meta-atoms in the x-axis and y-axis directions, respectively. The meta-atom spacing in the x-axis direction and y-axis direction is represented as $d_{x}^{\text{M}}$ and $d_{y}^{\text{M}}$, respectively. The spacing between metasurfaces is denoted as $d_{\text{layer}}$, while the spacing between the UPA and the SIM is denoted as $d_{\text{US}}$. Furthermore, for convenience, let $\mathcal{L}=\{1,2,\cdots,L\}$, $\mathcal{N}=\{1,2,\cdots,N\}$ and $\mathcal{M}=\{1,2,\cdots,M\}$ denote the set of metasurfaces, meta-atoms on each metasurface layer, and antennas on the UPA, respectively. And for the $n$-th meta-atom on each layer, we define 
\begin{equation}
\label{n_define}
n=n_x+(n_y-1)N_x,
\end{equation}
where $n_x=1,2,\cdots,N_x$ and $n_y=1,2,\cdots,N_y$ represent the x-axis and y-axis coordinates of the $n$-th meta-atom, respectively. Moreover, the same convention is adopted for the $m$-th antenna.

\begin{figure*}[!t]
\centering
\subfigure[]{
		\includegraphics[width=0.4\textwidth]{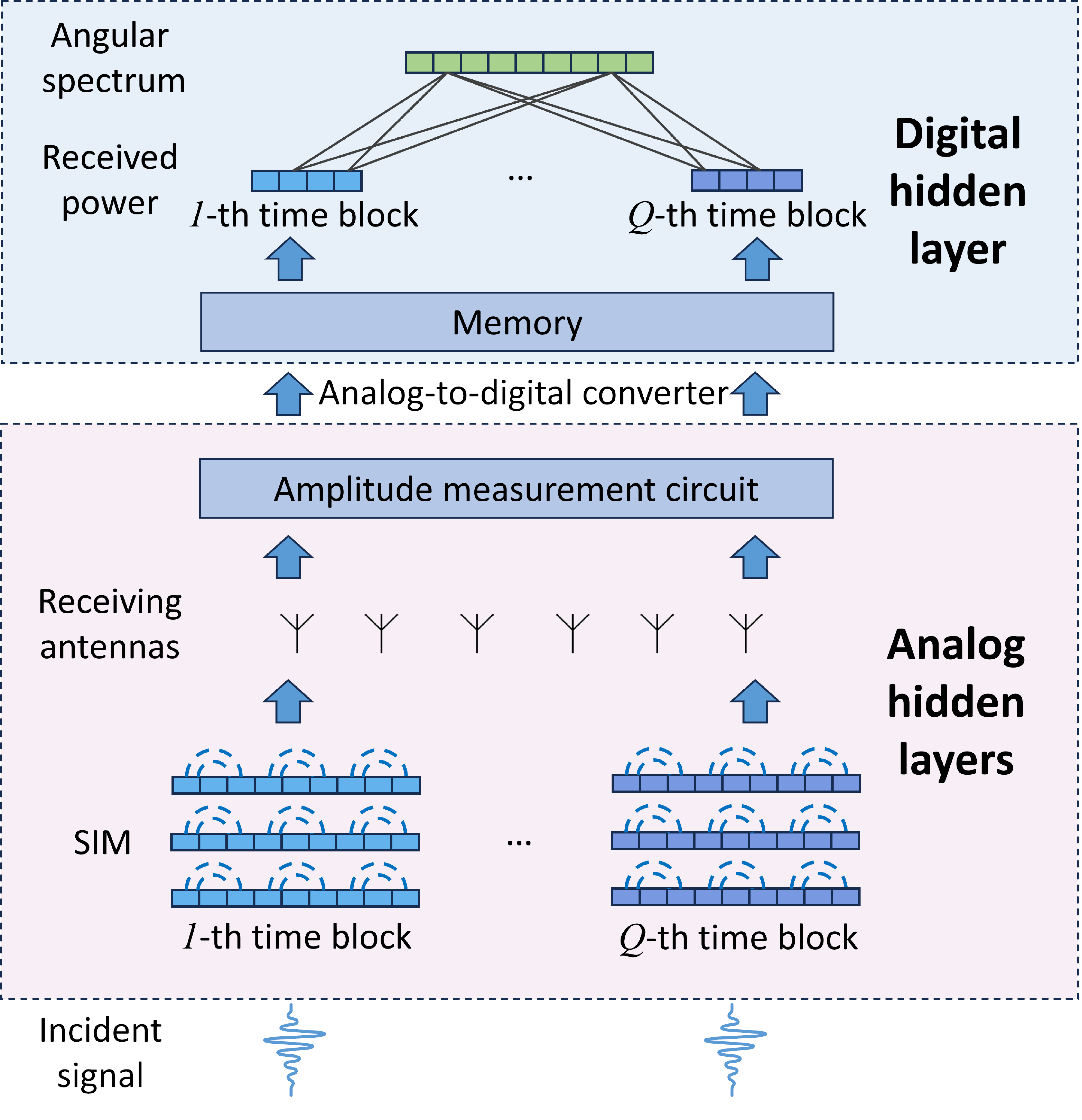}}
\subfigure[]{
		\includegraphics[width=0.45\textwidth]{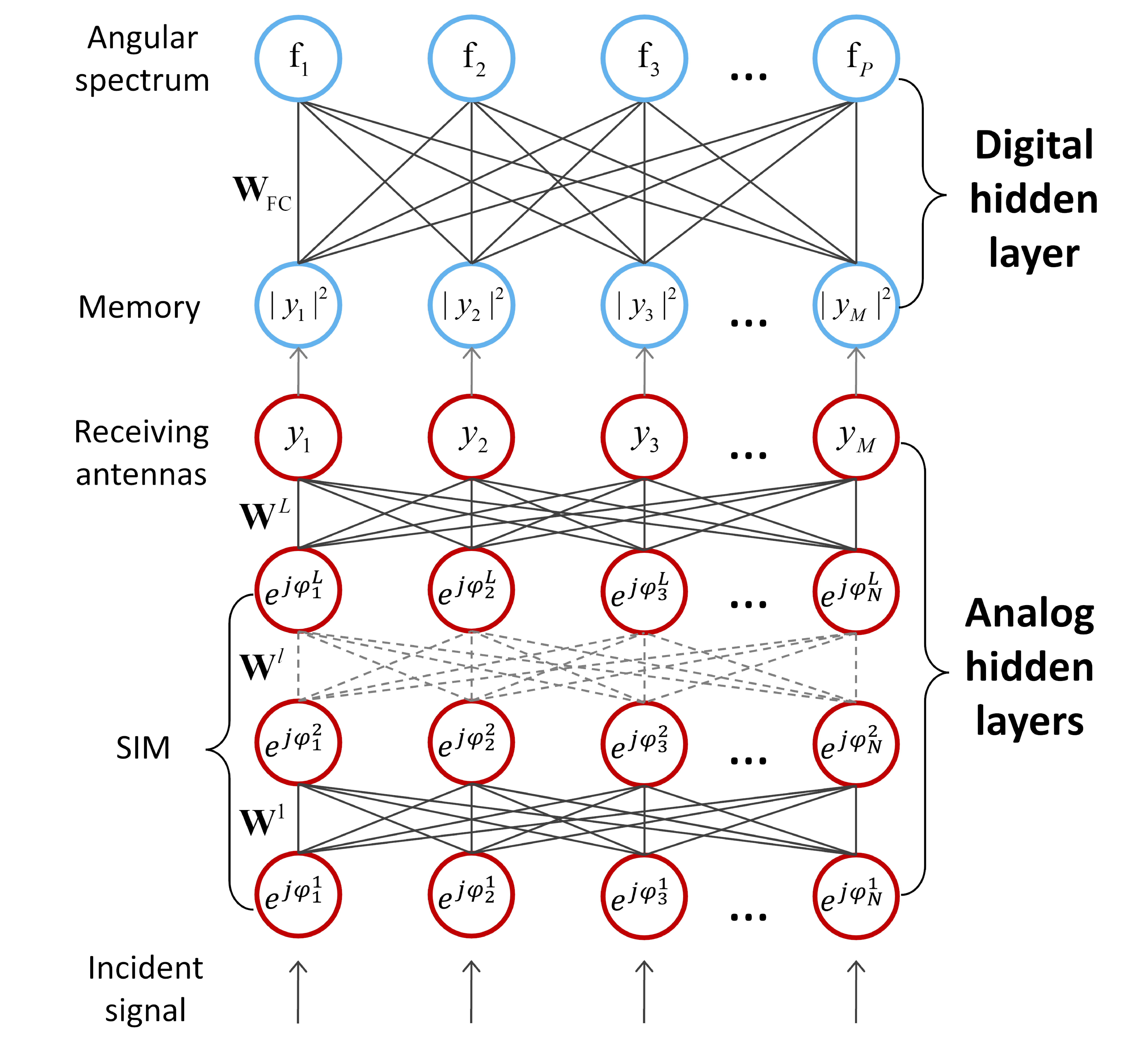}}
\caption{(a) The workflow of using an EMNN for DOA estimation. (b) The network architecture of the EMNN when $Q=1$.}
\label{fig_3}\vspace{-0.5cm}
\end{figure*}

Furthermore, we consider far-field narrowband incident signals radiating from $K$ distinct signal sources, denoted as ${\bf s}\in{\mathbb C}^{K \times 1}$. It is assumed that they are mutually independent with zero means, satisfying $\mathbb{E}[{\bf s}{\bf s}^H]=\text{diag}({\bf p}_s)$, where ${\bf p}_s=[p_1,\cdots,p_K]^T\in {\mathbb R}^{K \times 1}$ denotes incident signal power vector, $p_k$ denotes the power of the $k$-th signal. Furthermore, the incident azimuth and elevation angles of the $k$-th signal are represented as ${\theta^a_k} \in [0,2\pi)$ and ${\theta^e_k} \in [0,\pi/2)$, respectively. For convenience, we define ${\bm\theta}_k=[{\theta^a_k} ,{\theta^e_k} ]^T \in \mathbb{R}^{2\times1}$, and denote ${\bf\Theta}=\{{\bm\theta}_1 ,\cdots,{\bm\theta}_K \}$ as the set of incident angles. Therefore, the incident signal ${\bf s}^{\prime}\in{\mathbb C}^{N \times 1}$ at the first layer of SIM can be modeled as 
\begin{equation}
\label{s_model}
{\bf s}^{\prime}={\bf A}({\bf\Theta}){\bf s},
\end{equation}
where ${\bf A}({\bf\Theta})=[{\bf a}({\bm\theta}_1),\cdots,{\bf a}({\bm\theta}_K)]\in{\mathbb C}^{N \times K}$ is the array response matrix of the first layer of SIM associated with $K$ signal sources, with ${\bf a}({\bm\theta}_k)\in{\mathbb C}^{N \times 1}$ being the array response vector of the $k$-th signal. For a specific ${\bm\theta}$, the array response vector of the first layer of SIM is given by
\begin{equation}
\label{a_model}
{\bf a}({\bm\theta})={\bf g}(u_y)\otimes{\bf g}(u_x).
\end{equation}
In \eqref{a_model}, the vectors ${\bf g}(u_x)\in{\mathbb C}^{N_x \times 1}$ and ${\bf g}(u_y)\in{\mathbb C}^{N_y \times 1}$ are respectively represented as
\begin{subequations}
\begin{align}
{\bf g}(u_x)&=[1,e^{j u_x},\cdots,e^{j u_x(N_x-1)}]^T,
\\{\bf g}(u_y)&=[1,e^{j u_y},\cdots,e^{j u_y(N_y-1)}]^T,
\end{align}
\end{subequations}
where $u_x=\kappa d_x^{\text{M}}\sin{\theta^e}\cos{\theta^a}$, $u_y=\kappa d_y^{\text{M}}\sin{\theta^e}\sin{\theta^a}$, and $\kappa={2 \pi}/{\lambda}$ is the wavenumber.

In this paper, we aim to estimate the DOA parameters of $K$ signal sources by utilizing an EMNN. The DOA estimation problem is modeled as a multi-label classification task. Specifically, let $P$ denote the total number of classes to be classified and ${\mathcal{P}}=\{{\mathcal{C}}_1,{\mathcal{C}}_2,\cdots,{\mathcal{C}}_P\},{\mathcal{C}_{i}}\cap{\mathcal{C}_{j}}=\varnothing$ denote the designed DOA partition set. Moreover, let ${\bf f}\in{\mathbb R}^{P \times 1}$ denote the output of the EMNN. The aim is to train an EMNN to result in a higher value of $[{\bf f}]_p$ when the DOA parameter of an incident signal falls within ${\mathcal{C}}_p$.

\section{EMNN for DOA Estimation}
In this section, we introduce the structure of EMNN and the forward propagation model for inference. As shown in Fig. \ref{fig_3}, an EMNN is a cascade of metasurfaces as analog hidden layers and a digital hidden layer.

\subsection{Analog Hidden Layers}
As illustrated in Fig. \ref{fig_3}(a), when the signal passes through the analog hidden layers inside the SIM, it is equivalent to performing cascaded matrix operations in the EM domain. Therefore, the transfer function of SIM ${\bf B}({\bf{\Phi}})\in{\mathbb C}^{M \times N}$ can be expressed as
\begin{equation}
\label{G_model}
{\bf B}({\bf{\Phi}})={\bf W}^L{\bf\Lambda}^{L}{\bf W}^{L-1}\cdots{\bf\Lambda}^2{\bf W}^1{\bf\Lambda}^1,
\end{equation}
where ${\bf{\Phi}} =\{\phi^1_1,\cdots,\phi^L_N\}$ denotes the set of phase shifts in the SIM and $\phi^l_n \in (0,2\pi], \forall l\in\mathcal{L}, \forall n\in\mathcal{N}$ denotes the phase shift imposed by the $n$-th meta-atom on the $l$-th metasurface layer, ${\bf\Lambda}^l = {\text {diag}}(e^{j\phi^l_1},e^{j\phi^l_2},\cdots,e^{j\phi^l_N})\in{\mathbb C}^{N \times N}$ denotes the phase shift matrix induced by the $l$-th metasurface layer, and ${\bf W}^l\in{\mathbb C}^{N \times N},\ \forall l\ne L,\ l\in\mathcal{L}$ denotes the propagation matrix from the $l$-th metasurface layer to the $(l+1)$-th metasurface layer, where ${[\bf W}^l]_{n,n^{\prime}}$ represents the attenuation coefficient from the $n^{\prime}$-th meta-atom on the $l$-th metasurface layer to the $n$-th meta-atom on the $(l+1)$-th metasurface layer. Following the Rayleigh-Sommerfeld diffraction theory\cite{lin2018all}, ${[\bf W}^l]_{n,n^{\prime}}$ is given by
\begin{equation}
\label{w_model}
{[\bf W}^l]_{n,n^{\prime}}={\frac{S_{\text{ meta-atom}}d_{\text{layer}}}{({r^l_{n,n^{\prime}}})^2}}\left({\frac{1}{2\pi r^l_{n,n^{\prime}}}}-j{\frac{1}{\lambda}}\right) e^{\frac{j2\pi{r^l_{n,n^{\prime}}}}{\lambda}}, 
\end{equation}
where $S_{\text{meta-atom}}$ denotes the size of each meta-atom, $r^l_{n,n^{\prime}}$ denotes the corresponding propagation distance, and $\lambda$ denotes the radio wavelength. More specifically, $r^l_{n,n^{\prime}}$ can be expressed as
\begin{equation}
\label{r_model}
r^l_{n,n^{\prime}}=\sqrt{({D_x^{n,n^{\prime}}})^2+({D_y^{n,n^{\prime}}})^2+d_{\text{layer}}^2}, 
\end{equation}
where ${D_x^{n,n^{\prime}}}$ and ${D_y^{n,n^{\prime}}}$ represent the distances between two meta-atoms projected on the x-axis and y-axis, respectively. In the system setup considered, they can be calculated as
\begin{equation}
\label{D_model1}
{D_x^{n,n^{\prime}}}=|(n_x-n_x^{\prime})d_{x}^{\text{M}}|, \ {D_y^{n,n^{\prime}}}=|(n_y-n_y^{\prime})d_{y}^{\text{M}}|,
\end{equation}
respectively. Moreover, ${\bf W}^L\in{\mathbb C}^{M \times N}$ denotes the propagation matrix from the $L$-th layer to the UPA, and $[{\bf W}^L]_{m,n}$ can also be expressed by \eqref{w_model} and \eqref{r_model}, where the vertical distance $d_{\text{layer}}$ between layers is replaced by $d_{\text{US}}$. Besides, the distances between the $m$-th antenna of the UPA and the $n$-th meta-atom on the $L$-th layer projected in the x-axis and y-axis can be expressed as 
\begin{subequations}
\label{D_model2}
\begin{align}
{D_x^{m,n}}=|(m_x-\frac{M_x+1}{2})d^{\text{A}}_{x}-(n_x-\frac{N_x+1}{2})d_{x}^{\text{M}}|,
\\{D_y^{m,n}}=|(m_y-\frac{M_y+1}{2})d_{y}^{\text{A}}-(n_y-\frac{N_y+1}{2})d_{y}^{\text{M}}|,
\end{align}
\end{subequations}
respectively. 

Therefore, the signal received at the UPA ${\bf y}\in{\mathbb C}^{M\times 1}$ at a specific time $t$ can be modeled as 
 \begin{equation}
\label{y_model}
{\bf y}(t)={\bf B}({\bf{\Phi}}){\bf s}^{\prime}(t)+{\bf n}(t),
\end{equation}
where ${\bf n}(t)\in{\mathbb C}^{M \times 1}$ is the additive white Gaussian noise satisfying ${\bf n}(t)\in\mathcal{CN}({\bf0},\sigma_n^2{\bf I}_M)$, with $\sigma_n^2$ being the noise power.

After preprocessing in the EM domain, the power distribution across the receiving antenna array is measured using the amplitude measurement circuit to obtain the output of the analog hidden layers. Furthermore, to expand the scale of the analog hidden layers under practical hardware constraints, as shown in Fig. \ref{fig_3}(a), different SIM phase shift configurations are utilized in different time blocks, and the corresponding received signal power is recorded. Additionally, the received signal power under each SIM configuration is smoothed across multiple snapshots before inputting the fully connected layer.

Assuming $Q$ SIM configurations are utilized, let $\tilde{\bf y} = ({\bf y}_1^T,\cdots,{\bf y}_Q^T)^T\in {\mathbb C}^{MQ \times 1}$ denote the received signal under all $Q$ SIM configurations, where ${\bf y}_q$ denotes the received signal under the $q$-th SIM configuration. Specifically, considering the noise is uncorrelated with the incident signals, the output of the analog hidden layers can be calculated as
\begin{equation}
\label{ey1_model}
{\mathbb E}\left[|\tilde{\bf y}|^2\right]=\sum_{k=1}^{K}|\tilde{\bf B}{\bf a}({\bm\theta_k})|^2{ p}_k+\sigma_n^2{\bf 1},
\end{equation}
where $\tilde{\bf B} = [{\bf B}({\bf{\Phi}}_1)^T,\cdots,{\bf B}({\bf{\Phi}}_Q)^T]^T\in {\mathbb C}^{MQ \times N}$ denotes equivalent propagation matrix under all SIM configurations.

\textit{Remark 1}: Note that SIM extracts and transforms phase and amplitude distribution information of incident EM waves on the large-aperture metasurface into an amplitude pattern as they pass through the layered structure. Consequently, the system can estimate the DOA of targets by relying solely on the amplitude measurements. Existing DOA estimation methods, directly replacing amplitude-phase measurements with amplitude measurements \cite{SPL_2015_Kim_Non,TAES_2023_Wan_Target}, would result in severe ambiguity.

\textit{Remark 2}: Although the calculation of $\bf B$ in \eqref{G_model} involves a considerable amount of matrix operations, these computations naturally occur in the EM domain and are completed at the speed of light as the incident EM waves propagate through the SIM. This property forms the cornerstone of EMNN and incurs relatively low energy consumption.

\textit{Remark 3}: In this paper, we suppose that ${\bf W}^l,l\in \mathcal{L}$ are accurately obtainable. Although \eqref{w_model} to \eqref{D_model2} may not precisely capture the attenuation coefficients in practical scenarios due to hardware imperfections such as metasurface misalignment, the efficient backpropagation algorithm in \cite{liu2022programmable} can be utilized to calibrate these attenuation coefficients.

\subsection{Digital Hidden Layer}
After passing through a digital hidden layer, the output ${\bf f}$ of the EMNN can be represented as
\begin{equation}
\label{f_model}
{\bf f}={\bf W}_\text{FC} (\mathbb{E} \left[|\tilde{\bf y}|^2\right]-\sigma_n^2{\bf 1}),
\end{equation}
where ${\bf W}_\text{FC}\in {\mathbb R}^{P \times MQ}$ denotes the weight coefficients of the fully connected layer. In practice, the average power of the received signal in \eqref{f_model} is approximated by the mean of a limited number of samples. Upon collecting $S$ snapshots for each SIM configuration, the angular spectrum obtained is expressed as 
\begin{equation}
\label{f1_model}
\tilde{\bf f}={\bf W}_\text{FC}\left(\frac{1}{S}\sum_{s=1}^S|{\tilde{\bf y}_s}|^2-\sigma_n^2{\bf 1}\right).
\end{equation}
In \eqref{f1_model}, $\tilde{\bf y}_s=({\bf y}_1^T(t_{1,s}),\cdots,{\bf y}_Q^T(t_{Q,s}))^T\in {\mathbb C}^{MQ \times 1}$, where $t_{q,s}$ denotes the $s$-th snapshot slot under the $q$-th SIM configuration.

\textit{Remark 4}: In \eqref{f_model}, the noise power is used as the bias in the fully connected layer, which facilitates EMNN's adaptation to different SNR setups. Specifically, for an arbitrary bias ${\bf b}\in {\mathbb R}^{P \times 1}$, the obtained angular spectrum can be expressed as 
\begin{equation}
{\bf f}={\bf W}_\text{FC}\sum_{k=1}^{K}|\tilde{\bf B}{\bf a}({\bm\theta_k})|^2{ p}_k+\sigma_n^2{\bf W}_\text{FC}{\bf 1} + {\bf b}.
\end{equation}
When the bias ${\bf b}$ is fixed, and $\sigma_n^2{\bf W}_\text{FC}{\bf 1}\neq \xi {\bf 1}, \xi \in \mathbb{C}$, the peak location of $f$ will be affected by the weighted noise floor, even if the noise power is perfectly estimated. By contrast, employing \eqref{f_model} as the forward propagation model can eliminate this issue.

\textit{Remark 5}: Most traditional DOA estimation methods \cite{krim1996two,schmidt1986multiple,roy1989esprit,barabell1983improving,haardt1995unitary,papageorgiou2021deep,liu2018direction,elbir2020deepmusic,merkofer2022deep} require the measurement of the covariance matrix of the received signals ${\mathbb E}\left[{\bf y}{\bf y}^H\right]$, which necessitates complex amplitude-phase measurement circuits composed of down-conversion and I/Q separation. Additionally, smoothing over $S$ snapshots involves $4SM^2$ real multiplications and $(4S-2)M^2$ real additions. In contrast, the methods based on amplitude measurement only require measuring the received signal power distribution ${\mathbb E}\left[|{\bf y}|^2\right]$. This requires only low-complexity and energy-efficient amplitude measurement circuits, and smoothing involves $(S-1)M$ real additions.

\textit{Remark 6}: The CRB lower-bounds the covariance of any unbiased estimator \cite{kay1993fundamentals,stoica1989musiccrb}. For a parametric probability density function $f({\bf r};{\bm\theta})$ of an observation ${\bf r}$ with respect to the parameter vector ${\bm\theta}=[\theta^{a},\theta^{e}]^{T}$, the Fisher information matrix (FIM) ${\bf F}({\bm\theta})\in\mathbb{R}^{2\times 2}$ is defined as \cite{kay1993fundamentals}
\begin{equation}\label{fim_def}
{\bf F}({\bm\theta}) = {\mathbb E}\!\left\{ \nabla_{\bm\theta}\ln f({\bf r};{\bm\theta})\, \nabla_{\bm\theta}^{T}\ln f({\bf r};{\bm\theta}) \right\},
\end{equation}
based on which the angular root-mean-square error (RMSE) $\sqrt{{\mathbb E}[\|\hat{\bm\theta}-{\bm\theta}\|_{F}^{2}]}$ is lower-bounded by \cite{kay1993fundamentals}
\begin{equation}\label{crb_euclid_def}
\operatorname{CRB}_{\rm E}({\bm\theta}) = \sqrt{\operatorname{tr}\!\left({\bf F}^{-1}({\bm\theta})\right)}.
\end{equation}

Consider the single-source, single-snapshot, and constant-modulus case, and let ${\bf B}_{q}={\bf B}({\bm\Phi}_{q})$ denote the $q$-th block of $\tilde{\bf B}$ in \eqref{ey1_model}. The angular derivatives of ${\bf a}({\bm\theta})$ is given by
\begin{subequations}\label{da_dtheta}
\begin{align}
\frac{\partial{\bf a}({\bm\theta})}{\partial\theta^{a}}
 &=j\!\left[\dot u^{a}_{x}({\bf I}_{N_{y}}\!\otimes\!{\bm\Omega}_{x})
 +\dot u^{a}_{y}({\bm\Omega}_{y}\!\otimes\!{\bf I}_{N_{x}})\right]
 {\bf a}({\bm\theta}),\\
\frac{\partial{\bf a}({\bm\theta})}{\partial\theta^{e}}
 &=j\!\left[\dot u^{e}_{x}({\bf I}_{N_{y}}\!\otimes\!{\bm\Omega}_{x})
 +\dot u^{e}_{y}({\bm\Omega}_{y}\!\otimes\!{\bf I}_{N_{x}})\right]
 {\bf a}({\bm\theta}),
\end{align}
\end{subequations}
where ${\bm\Omega}_{x}={\text{diag}}(0,1,\ldots,N_{x}-1)$, ${\bm\Omega}_{y}={\text{diag}}(0,1,\ldots,N_{y}-1)$, and the phase-gradient scalars are
\begin{subequations}\label{u_dot}
\begin{align}
\dot u^{a}_{x}&=-\kappa d_{x}^{\text{M}}\sin\theta^{e}\sin\theta^{a},\ \
\dot u^{a}_{y}= \kappa d_{y}^{\text{M}}\sin\theta^{e}\cos\theta^{a},\\
\dot u^{e}_{x}&= \kappa d_{x}^{\text{M}}\cos\theta^{e}\cos\theta^{a},\ \
\dot u^{e}_{y}= \kappa d_{y}^{\text{M}}\cos\theta^{e}\sin\theta^{a}.
\end{align}
\end{subequations}

For the $m$-th antenna at the $q$-th SIM configuration, further define the effective steering scalar and its angular derivatives as
\begin{subequations}\label{scalars}
\begin{align}
\alpha_{m,q}({\bm\theta})
 &= [{\bf B}_{q}{\bf a}({\bm\theta})]_{m},\\
\beta^{a}_{m,q}({\bm\theta})
 &= \left[{\bf B}_{q}\,\tfrac{\partial{\bf a}({\bm\theta})}{\partial\theta^{a}}\right]_{m},\\
\beta^{e}_{m,q}({\bm\theta})
 &= \left[{\bf B}_{q}\,\tfrac{\partial{\bf a}({\bm\theta})}{\partial\theta^{e}}\right]_{m}.
\end{align}
\end{subequations}

\begin{figure*}[!t]
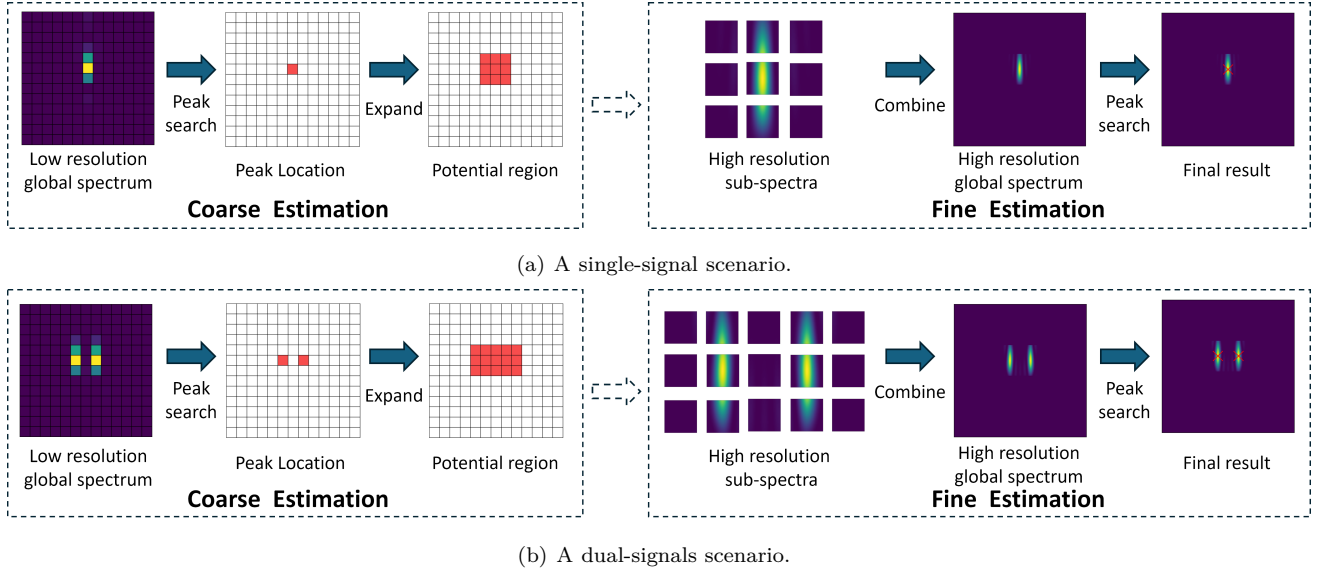

\centering
\subfigure[A single-signal scenario.]{
		\includegraphics[width=0.95\linewidth]{Fig/protocol1.eps}}
\subfigure[A dual-signals scenario.]{
		\includegraphics[width=0.95\linewidth]{Fig/protocol2.eps}}
\caption{Schematic diagram of the proposed hierarchical DOA estimation framework.}
\label{fig_4}\vspace{-0.5cm}
\end{figure*}

Stacking the real and imaginary parts of the per-antenna cross-correlations yields the in-phase and quadrature gradient vectors
\begin{subequations}\label{eta_and_etaperp}
\begin{align}
{\bm\eta}_{m,q}({\bm\theta})
 &=2\rho\!\begin{bmatrix}
 \operatorname{real}\!\left\{\alpha_{m,q}^{*}\beta^{a}_{m,q}\right\}\\[2pt]
 \operatorname{real}\!\left\{\alpha_{m,q}^{*}\beta^{e}_{m,q}\right\}
 \end{bmatrix},\\
{\bm\eta}^{\perp}_{m,q}({\bm\theta})
 &=2\rho\!\begin{bmatrix}
 \operatorname{imag}\!\left\{\alpha_{m,q}^{*}\beta^{a}_{m,q}\right\}\\[2pt]
 \operatorname{imag}\!\left\{\alpha_{m,q}^{*}\beta^{e}_{m,q}\right\}
 \end{bmatrix},
\end{align}
\end{subequations}
where we have $\rho= p_{1}/\sigma_{n}^{2}$, $\gamma_{m,q}({\bm\theta})=\rho|\alpha_{m,q}({\bm\theta})|^{2}$.

The phase-aware coherent FIM is obtained from the classical array CRB \cite{kay1993fundamentals,stoica1989musiccrb} by substituting the conventional steering vector ${\bf a}({\bm\theta})$ with the SIM-filtered effective steering vector ${\bf B}_{q}{\bf a}({\bm\theta})$, which can be recast as
\begin{equation}\label{appendix_coh_v3}
{\bf F}_{\rm coh}({\bm\theta})
= \sum_{q=1}^{Q}\sum_{m=1}^{M}
 \frac{{\bm\eta}_{m,q}{\bm\eta}_{m,q}^{T}
 +{\bm\eta}^{\perp}_{m,q}({\bm\eta}^{\perp}_{m,q})^{T}}
 {2\gamma_{m,q}({\bm\theta})}.
\end{equation}

Under the amplitude-only observation, the scalar Fisher information can be expressed as
\begin{equation}\label{J1_def}
{\mathcal J}_{1}(\gamma)
=\!\int_{0}^{\infty}\!\!\!\bigl[\sqrt{z/\gamma}\,\tfrac{I_{1}(2\sqrt{\gamma z})}{I_{0}(2\sqrt{\gamma z})}-1\bigr]^{2}\!e^{-(z+\gamma)}I_{0}(2\sqrt{\gamma z})\,{\rm d}z,
\end{equation}
where $I_{0}(\cdot)$, $I_{1}(\cdot)$ and $z_{m,q}=|y_{m,q}|^{2}/\sigma_{n}^{2}$ denote the modified Bessel functions of the first kind of orders zero and one, and the normalized power, respectively. Furthermore, the amplitude-only FIM is given by
\begin{equation}\label{crb_power_fim_remark_v3}
{\bf F}_{\rm pow}({\bm\theta})
= \sum_{q=1}^{Q}\sum_{m=1}^{M}
 {\mathcal J}_{1}\!\big(\gamma_{m,q}({\bm\theta})\big)\,
 {\bm\eta}_{m,q}({\bm\theta})\,{\bm\eta}_{m,q}^{T}({\bm\theta}).
\end{equation}
The detailed derivation of \eqref{crb_power_fim_remark_v3} is provided in Appendix~A.

\section{Hierarchical DOA Estimation Based on Angular Spectrum Subregions}
Although the proposed EMNN can be directly applied to the global DOA estimation, its requirement for a large number of SIM configurations in high-resolution scenarios limits its applicability. In this section, to reduce the computational complexity and the number of observation snapshots required for DOA estimation, a hierarchical DOA estimation method is proposed. Specifically, Section IV-A introduces the framework of the proposed hierarchical protocol. In Section IV-B, we elaborate on the grid partitioning method. Section IV-C introduces the corresponding training methodology of EMNN, while Section IV-D offers a comprehensive summary of the entire algorithm.

\subsection{Hierarchical DOA Estimation Framework} 
As shown in Fig. \ref{fig_4}, the entire DOA estimation process is divided into two stages: coarse estimation and fine estimation. In the coarse estimation stage, an EMNN is trained to generate the low-resolution global angular spectrum, based on which a potential region containing the DOA of incident signals is determined for fine estimation. In the fine estimation stage, multiple EMNNs are used to generate a high-resolution angular spectrum within the potential region, and the final DOA estimate is obtained by searching across this spectrum.

Next, we elaborate on the procedures involved in these two stages. 
\begin{itemize}
\item \textbf{Coarse estimation}: The entire DOA range is uniformly divided into $G$ subregions $\{\mathcal{G}_g\}$, from which a potential region is tailored for conducting the fine estimation. First, a global low-resolution spectrum is generated using an EMNN, which is pre-trained for recognizing these $P=G$ subregions. Then, the spectrum peaks are identified, based on which the potential region is constructed by incorporating peaks and their surroundings. The expansion is due to the fact that the DOA may be located at the grid boundary. Specifically, let $\mathcal{G}_{g_k}$ denote the subregion containing the peak found in the coarse estimation phase for the $k$-th target. The expanded potential region $\mathcal{R}_{k}$ for the fine estimation consists of $\mathcal{G}_{g_k}$ and its eight neighboring subregions surrounding $\mathcal{G}_{g_k}$. Whereas, for $\mathcal{G}_{g_k}$ located at the maximum and minimum boundaries of the elevation angle, there are five surrounding subregions. Assuming that $K$ peaks can be identified, $\mathcal{R}=\bigcup_{k=1}^{K}\mathcal{R}_{k}$ is denoted as the potential region identified for fine estimation.
\item \textbf{Fine estimation}: In the fine estimation phase, each $\mathcal{G}_g$ is uniformly divided into $V$ DOA regions $\{{\mathcal{A}}_{g,v}\}$, and an EMNN is utilized to generate the corresponding sub-spectrum. In contrast to the coarse estimation, the EMNNs in the fine estimation stage also need to identify the region outside of $\mathcal{G}_g$, so it is required to recognize $P=V+1$ subregions. By generating and combining all sub-spectra for the subregions within the potential region $\mathcal{R}$, a high-resolution global spectrum is obtained. Ultimately, applying peak searching on this spectrum yields the DOA estimate with fine granularity.
\end{itemize}

\textit{Remark 7}: The hierarchical estimation method improves estimation accuracy by progressively narrowing the range of the generated angular spectrum. This fully enhances the inference capability of the EMNN, which in turn lowers the requirements for the number of SIM configurations and observation snapshots required. Moreover, the digital computation burden imposed on the fully connected layer is also mitigated. Taking a single incident signal as an example, the computational complexity for generating the angular spectrum is directly proportional to its output size. Therefore, the condition for the hierarchical estimation method to achieve lower computational complexity in angular spectrum generation compared to the exclusive EMNN approach is given by the $G + 9V < GV$. This inequality is equivalent to$(G - 9)(V - 1) > 9$, which can be readily satisfied when the grid resolutions, i.e., $G$ and $V$, are sufficiently high.

\subsection{Grid Partitioning } 
In this subsection, we will elucidate the grid partitioning of the angular spectrum for implementing the proposed hierarchical DOA estimation framework. To facilitate this, we define three distinct sets: $\{\mathcal{G}_g\}$, $\{\mathcal{R}_k\}$ and $\{{\mathcal{A}}_{g,v}\}$.
\begin{itemize}
 \item \textbf{$\{\mathcal{G}_g\}$} corresponds to the global angular spectrum generated for coarse estimation. It is obtained by uniformly dividing the entire DOA region into $G=G_a\times G_e$ parts based on azimuth and elevation angles. Specifically, for a given subregion index $g$, we define:
\begin{equation}
\label{g_define}
g=g^a+(g^e-1)G_a,
\end{equation}
where $g^a=1,2,\cdots,G_a$ and $g^e=1,2,\cdots,G_e$. Therefore, ${\mathcal{G}}_g$ can be expressed as $\mathcal{G}_g=\{{\bm\theta}|\theta^a\in[\theta^{a,\text{start}}_g,\theta^{a,\text{start}}_g+E^{a,\text{coarse}}),\theta^e\in[\theta^{e,\text{start}}_g,\theta^{e,\text{start}}_g+E^{e,\text{coarse}})\}$, where we have
\begin{subequations}
\begin{align}
&\theta^{a,\text{start}}_g=(g^a-1)E^{a,\text{coarse}},\\
&\theta^{e,\text{start}}_g=(g^e-1)E^{e,\text{coarse}},\\
&E^{a,\text{coarse}}=\frac{2\pi}{G_a},E^{e,\text{coarse}}=\frac{\pi}{2G_e}.
\end{align}
\end{subequations}
\item \textbf{$\{\mathcal{R}_k\}$} represent the enlarged potential region according to $\{\mathcal{G}_{g_k}\}$ obtained in the coarse estimation stage. For a given ${\mathcal{G}}_{g_k}$, the enlarged set $\mathcal{R}_k$ includes $\mathcal{G}_{g_k}$ and its surrounding subregions. Considering the periodicity of the azimuth angle, it can be represented as 
\begin{subequations}
\begin{align}
&\mathcal{R}_k=\left\{\mathcal{G}_{g}|l^{e}_{g,g_k}\leq 1 ,\ l^{a}_{g,g_k} \leq 1 \right\},\\
&l^{e}_{g,g_k}=|g^e-g^e_k|,\\
&l^{a}_{g,g_k}=\min\{|g^a-g^a_k|,G_a-|g^a-g^a_k|\}.
\end{align}
\end{subequations}
\item \textbf{$\{{\mathcal{A}}_{g,v}\}$} represent the local angular spectrum produced for fine estimation. Similarly, for each $g$, $\{{\mathcal{A}}_{g,v}\}$ is derived by uniformly dividing $\mathcal{G}_g$ into $V=V_a\times V_e$ parts. After rewriting $v$ in a similar form as \eqref{g_define}, ${\mathcal{A}}_{g,v}$ can be expressed as ${\mathcal{A}}_{g,v}=\{{\bm\theta}|\theta^a\in[\theta^{a,\text{start}}_{g,v},\theta^{a,\text{start}}_{g,v}+E^{a,\text{fine}}),\theta^e\in[\theta^{e,\text{start}}_{g,v},\theta^{e,\text{start}}_{g,v}+E^{e,\text{fine}})\}$, where 
\begin{subequations}
\begin{align}
&\theta^{a,\text{start}}_{g,v}=\theta^{a,\text{start}}_{g}+(v^a-1)E^{a,\text{fine}},\\
&\theta^{e,\text{start}}_{g,v}=\theta^{e,\text{start}}_{g}+(v^e-1)E^{e,\text{fine}},\\
&E^{a,\text{fine}}=\frac{2\pi}{G_aV_a},E^{e,\text{fine}}=\frac{\pi}{2G_eV_e}.
\end{align}
\end{subequations}
\end{itemize}

\subsection{Training Procedure}
In this subsection, we will elaborate on the training procedures of the EMNNs used for the hierarchical DOA estimation method. This includes the generation of the training set, comprising the methods for sampling and labeling, as well as the forward propagation model, and the normalization of weight coefficients for sub-spectra concatenation.

\subsubsection {Training Data Generation}
For the EMNN used for coarse estimation, we first randomly generate a single-signal DOA dataset $\{\hat{\bm\theta}\}$, which is composed of a large number of randomly generated samples from the entire DOA region. Furthermore, each sample in multiple-signal training set $\{\hat{\bf\Theta}\}$ is generated by combining $K$ samples extracted randomly from $\{\hat{\bm\theta}\}$. For the EMNN used in fine estimation corresponding to a certain ${\mathcal{G}}_{g}$, a similar method is used to construct the training set. However, in the fine estimation stage, the single-signal training set $\{\hat{\bm\theta}\}$ needs to include samples from both ${\mathcal{G}}_{g}$ and other subregions outside ${\mathcal{G}}_{g}$ to suppress the influence of signals from those subregions. The higher the proportion of samples from ${\mathcal{G}}_{g}$, the stronger EMNN's ability to identify signals within that region, which, however, weakens its ability to suppress signals from other subregions. Assuming that ${\mathcal{R}}_{k}$ represents the expanded potential region when the peak is located in ${\mathcal{G}}_{g}$, to minimize interference from signals in the surrounding subregions ${\mathcal{R}}_{k}-{\mathcal{G}}_{g}$, the ratio of the number of samples in $\{\hat{\bm\theta}\}$ within ${\mathcal{G}}_{g}$, ${\mathcal{R}}_{k}-{\mathcal{G}}_{g}$, and other subregions is set to an empirical value of 1:1:1.

\begin{table}[t]
\renewcommand{\arraystretch}{1.25}
\centering
\caption{Examples of Labeling Method}
\begin{tabular}{c||c}
\hline
DOA distribution &
Label \\ \hline
\parbox[c][0.25\linewidth][c]{0.25\linewidth}{\centering\includegraphics[width={1\linewidth}]{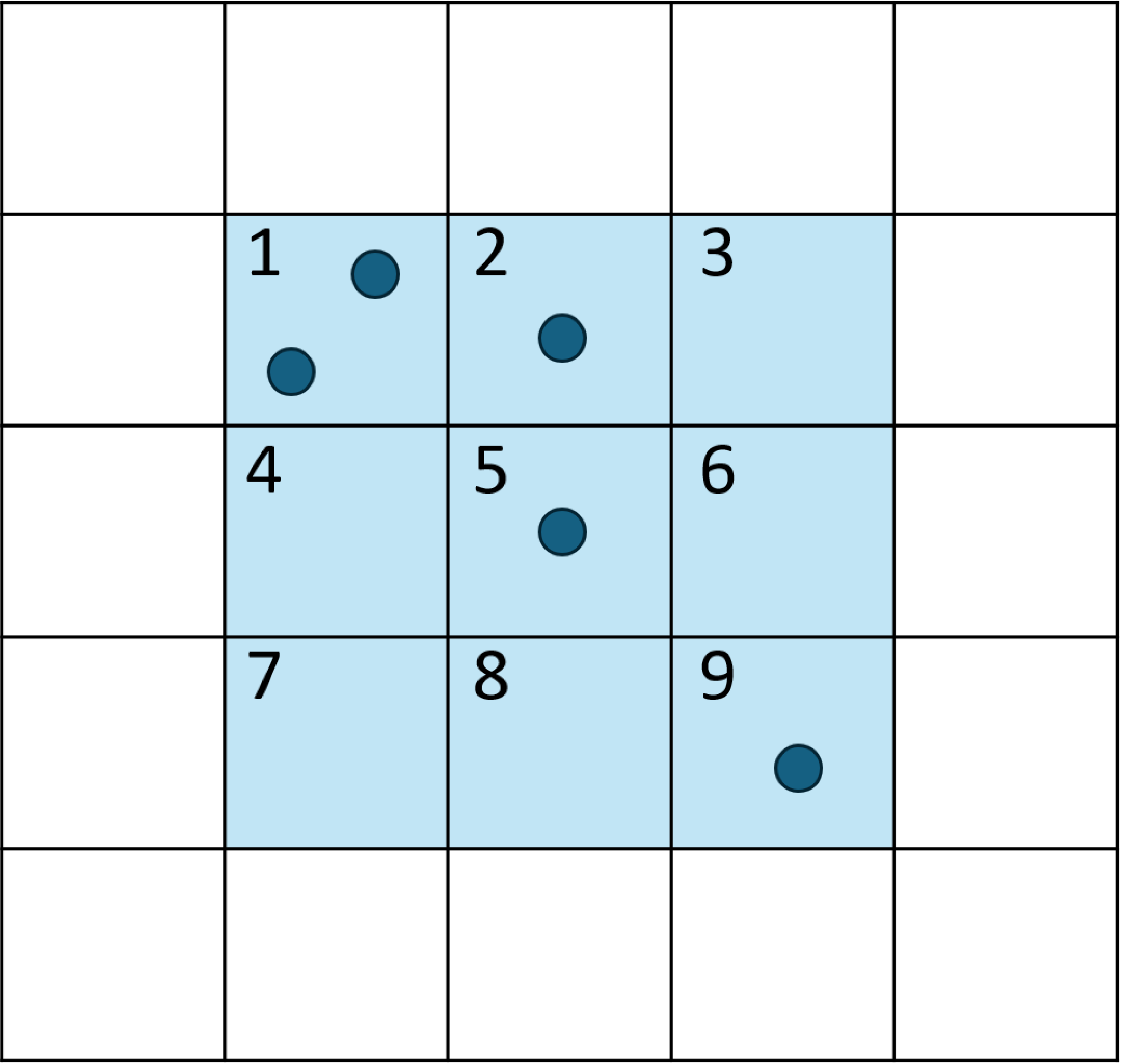}} &
\parbox[c][0.25\linewidth][c]{0.6\linewidth}{\centering
 \fontsize{11}{13}\selectfont $\hat{\bf q} =[\frac{2}{5}, \frac{1}{5}, 0, 0,\frac{1}{5},0,0,0,\frac{1}{5},0]^T$
}\\ \hline
\parbox[c][0.25\linewidth][c]{0.25\linewidth}{\centering\includegraphics[width={1\linewidth}]{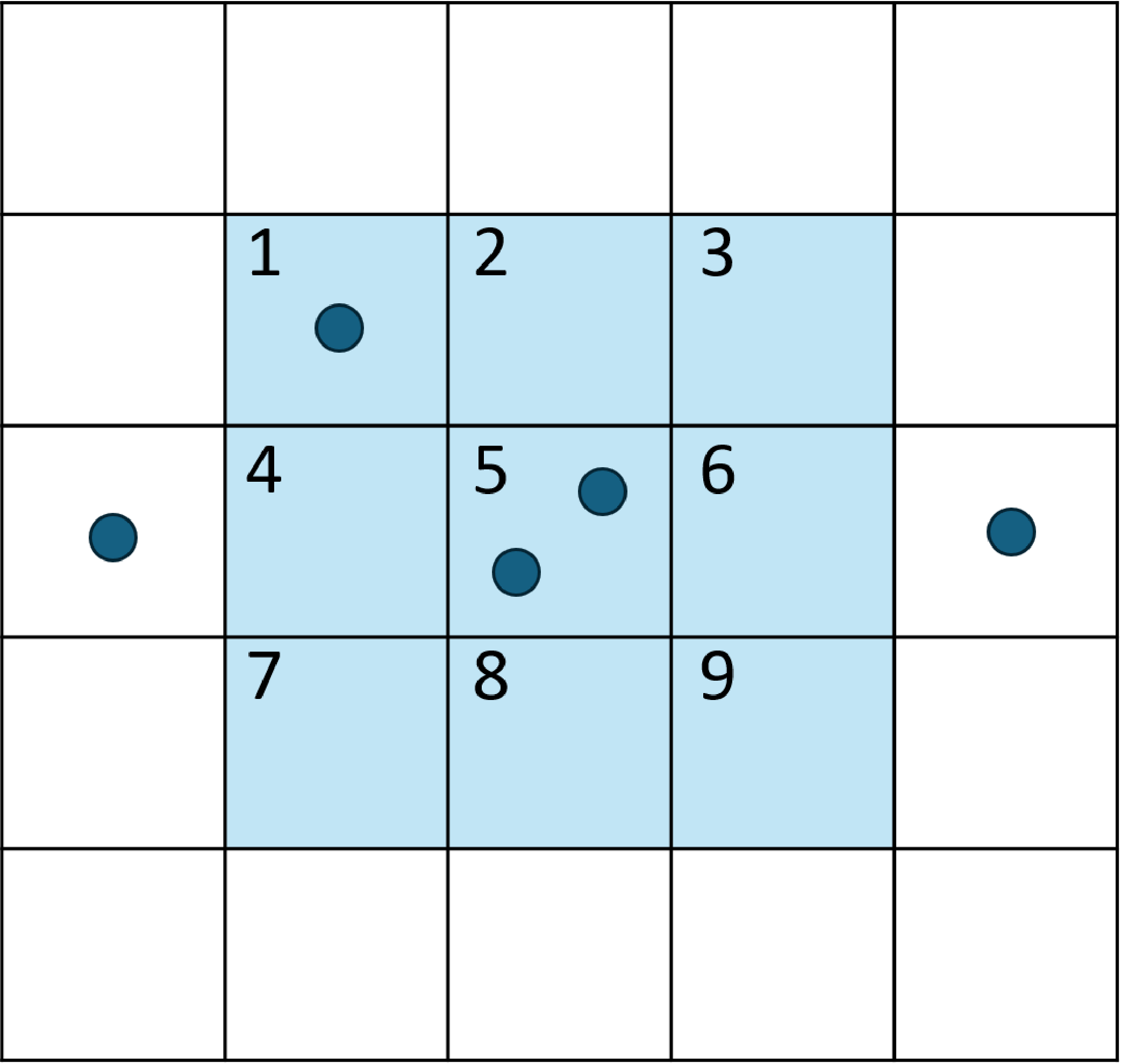}} &
\parbox[c][0.25\linewidth][c]{0.6\linewidth}{\centering
 \fontsize{11}{13}\selectfont $\hat{\bf q} = [\frac{1}{5}, 0, 0, 0,\frac{2}{5},0,0,0,0,\frac{2}{5}]^T$
}\\ \hline
\parbox[c][0.25\linewidth][c]{0.25\linewidth}{\centering\includegraphics[width={1\linewidth}]{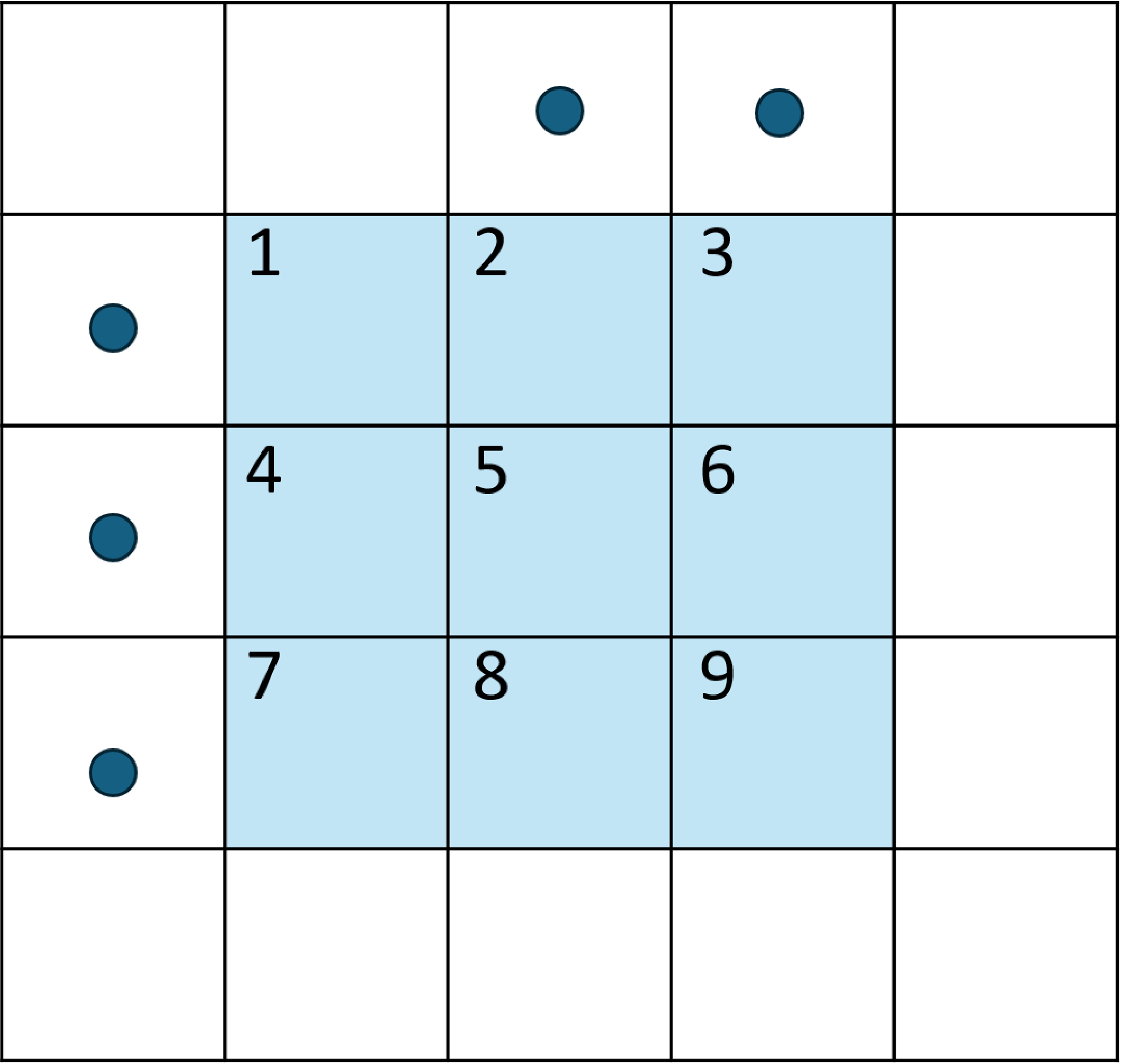}} &
\parbox[c][0.25\linewidth][c]{0.6\linewidth}{\centering
 \fontsize{11}{13}\selectfont $\hat{\bf q} =[0,0,0,0,0,0,0,0,0,1]^T$
}\\ \hline
\end{tabular}\vspace{-0.5cm}
\end{table}

\subsubsection{Data Labeling}
The training data needs to be labeled to map the target to its corresponding spectrum. 
Specifically, for a given sample $\hat{\bf\Theta}$ for training the EMNN to generate the spectrum in ${\mathcal{P}}$, let $K_p$ denote the number of $\hat{\bm\theta}_k$ fall within the $p$-th subregion $\mathcal{C}_p$. Moreover, $p_{\text{trn}}$ is denoted as the power of all incident signals during the training process, i.e., $p_k=p_{\text{trn}},k=1,\cdots,K$, and the target output probability distribution $\hat{\bf q}\in{\mathbb R}^{P \times 1}$ is used as the label for calculating the loss function. Specifically, the $p$-th element of $\hat{\bf q}$ can be represented as
\begin{equation}
\label{p}
[\hat{\bf q}]_p = \frac{K_p}{K}, p=1,\cdots,P.
\end{equation}

Specifically, for the EMNN used for coarse estimation, it needs to distinguish $\{{\mathcal{G}}_{g}\}$, which means $P=G$ and ${\mathcal{C}}_{p}={\mathcal{G}}_{p}, p=1,\cdots,G$. In contrast, the EMNN used for fine estimation corresponding to a given ${\mathcal{G}}_{g}$ encounters an open-set recognition problem. Therefore, it needs to distinguish subregions including $\{{\mathcal{A}}_{g,v}\}\subset {\mathcal{G}}_{g}$ and the DOA region outsides ${\mathcal{G}}_{g}$, which means $P=V+1$, ${\mathcal{C}}_{p}={\mathcal{A}}_{g,p}$, for $p=1,\cdots,V$ and ${\mathcal{C}}_{p}={\mathcal{O}}_{g}$ for $p=V + 1$, where ${\mathcal{O}}_{g}$ denotes the DOA region outsides ${\mathcal{G}}_{g}$. To illustrate, we present three examples of DOA samples and their corresponding labels for training EMNNs used for fine estimation in \textbf{Table IV}. On the left side of the \textbf{Table IV}, each square corresponds to a high-resolution subregion $A_{g,v}$, the shaded area indicates the coarse-resolution subregion ${\mathcal{G}}_g$, the number of blue dots, i.e., the targets, in each shaded grid defines the corresponding value in $\tilde{\bf q}$, according to the convention in \eqref{p}.

\subsubsection{Forward Propagation Model}
In the inference phase, the forward propagation model directly evaluates the output value for identifying the index of the peak. By contrast, in the training stage, an additional softmax layer is incorporated to expedite the convergence speed of the loss function. Additionally, to enhance the robustness of the EMNNs, an extra perturbation ${\bf n}_p\in{\mathbb C}^{MQ \times 1}\in\mathcal{CN}({\bf0},\sigma_p^2{\bf I}_{MQ})$ with power $\sigma^2_p$ is introduced during the training process. Therefore, based on \eqref{ey1_model} and \eqref{f_model}, the forward propagation model during training is given by 
\begin{equation}
\label{ez_model}
\hat{\bf z}({\bf\Theta})=\text{softmax}({\bf W}_\text{FC}(\sum_{k=1}^{K}|\tilde{\bf B}{\bf a}({\bm\theta_k})|^2{ p}_k+|{\bf n}_p|^2)).
\end{equation}

Moreover, the cross-entropy function is used as the loss function to train the EMNNs. Specifically, assuming the batch size is $D$, let $\hat{\bf z}_d,d=1,\cdots,D$ represent the output of the EMNN and $\hat{\bf q}_d,d=1,\cdots,D$ represent the corresponding probability distribution label, the loss function is defined as 
\begin{equation}
\label{L_model}
\text{Loss}=-\frac{1}{D}\sum^D_{d=1}\sum^P_{p=1}[{\hat{\bf z}}_d]_p\text{log}[\hat{\bf q}_d]_p.
\end{equation}

\textit{Remark 8}: It is noted that the perturbation term $|{\bf n}_p|^2$ in \eqref{ez_model} characterizes the error by using \eqref{f1_model} to approximate \eqref{f_model}, which contributes to enhancing the robustness of the EMNN against the measurement error. In each training epoch, ${\bf n}_p$ will be randomly regenerated.

\subsubsection {Normalization of Weight Coefficients}
As described in Section II, the expected peak power in the high-resolution spectrum, formed by splicing the sub-spectra in fine estimation, should be proportional to the incident signal power. Therefore, the weights of the fully connected layer ${\bf W}_\text{FC}$ of each EMNN used for fine estimation need to be normalized such that the peak heights of the spectrum generated by unit power signals incident from all directions are close to 1. Specifically, for an EMNN corresponding to certain $\mathcal{G}_g$, we extract $U$ samples $\hat{\bm\theta}_{u}, u=1,\cdots,U$ sampled from $\mathcal{G}_g$ and use the average of the peaks generated by these samples for normalization. Therefore, the normalization process can be represented as 
\begin{equation}
\label{normalization_define}
 {\bf W}_\text{FC} \leftarrow \frac{U}{\sum_{u=1}^{U}\max\{{\bf W}_\text{FC}|\tilde{\bf B}{\bf a}(\hat{\bm\theta}_{u})|^2\}}{\bf W}_\text{FC}.
\end{equation}
Furthermore, in the inference process for fine estimation, the last column of ${\bf W}_\text{FC}$ is removed since it corresponds to the region outside the subregion ${\mathcal{G}}_{g}$.

\subsection{Overall Algorithm} 
To elaborate, the proposed hierarchical DOA method based on EMNNs is summarized in \textbf{Algorithm 1}. It is noted that the execution of \textbf{Algorithm 1} hinges on the pre-training of all EMNNs, as well as the prior regarding the number of incident signals $K$ and the noise power $\sigma_n^2$.

\begin{algorithm}[t]
\caption{Hierarchical DOA Estimation Based on Angular Spectrum Subregions.}
\begin{algorithmic}[1]
\REQUIRE $K$, $S$, $\sigma_n^2$, all EMNN parameters.
\ENSURE ${\mathcal{A}}_{g_k,v_k}$ corresponding to ${\bm \theta}_k$.
\STATE Initialize the potential region to $\mathcal{R}=\varnothing$.
\STATE \underline {\textbf{Stage} 1. Coarse estimation:}
\STATE Use EMNN to obtain ${\bf f}^{\text{coarse}}\in{\mathbb R}^{G_aG_e\times1}$, and reshape ${\bf f}^{\text{coarse}}$ to obtain the global angular spectrum ${\bf F}^{\text{coarse}}\in{\mathbb R}^{G_a\times G_e}$. 
\STATE Perform a 2D spectral peak search on ${\bf F}^{\text{coarse}}$, extract the positions of the top $K$ peaks and the corresponding $\mathcal{G}_{g_k},k=1,\cdots,K$. 
\FOR{$k=1:K$}
\STATE Obtain $\mathcal{R}_{k}$ from $\mathcal{G}_{g_k}$.
\STATE Enlarge potential region: $\mathcal{R}=\mathcal{R}\bigcup\mathcal{R}_{k}$.
\ENDFOR 
\STATE Assuming $\mathcal{R}$ consists of $R$ subregions, the elements in $\mathcal{R}$ are denoted as $\mathcal{G}_{\delta_{r}},r=1,\cdots,R$, where $\{\delta_r\}$ denotes the indices corresponding to the subregions within $\mathcal{R}$.
\STATE \underline {\textbf{Stage} 2. Fine estimation:}
\FOR{$r=1:R$}
\STATE Use EMNN corresponding to $\mathcal{G}_{\delta_r}$ to obtain ${\bf f}^{\text{fine}}_{r}\in{\mathbb R}^{V_aV_e\times1}$, and reshape ${\bf f}^{\text{fine}}_{r}$ to obtain the sub-spectrum ${\bf F}^{\text{fine}}_{r}\in{\mathbb R}^{V_a\times V_e}$. 
\STATE Assign ${\bf F}^{\text{fine}}_{r}$ to the positions corresponding to $\mathcal{G}_{\delta_r}$ in angular spectrum for fine estimation ${\bf A} \in {\mathbb{R}}^{G_aV_a\times G_eV_e}$.
\ENDFOR 
\STATE Perform a 2D spectral peak search on ${\bf A}$ within $\mathcal{R}$, extract the positions of the top $K$ peaks, and obtain the final estimation result ${\mathcal{A}}_{g_k,v_k}$ corresponding to ${\bm \theta}_k$. 
\end{algorithmic}
\label{alg1}
\end{algorithm}

\section{Numerical Simulation}
In this section, we present the numerical results to evaluate the performance of the proposed DOA estimation method based on the EMNN.

\subsection{Simulation Setup and Performance Metrics} 
We consider a SIM integrated with the UAV. The number of metasurfaces, the number of meta-atoms per metasurface, and the number of receiving antennas are set to $L = 5$ and $N = 1600$, $M = 36$, respectively, with $N_x = N_y = 40$ and $M_x = M_y = 6$. The side length and size of the meta-atoms, spacing between adjacent receiving antennas, spacing between adjacent metasurfaces, spacing between metasurfaces and receiving antenna array are set to $d^{\text{M}}_x=d^{\text{M}}_y=\lambda /2$, $S_{\text{ meta-atom}} = \lambda^2/4$, $d_x^{\text{A}}=d_y^{\text{A}}=\lambda$, $d_{\text{layer}}=d_\text{US}=5\lambda$, respectively. Moreover, it is assumed that the incident signals $s_k,k =1,\cdots,K$ are constant-modulus signals with uniformly distributed and mutually independent phases, and the power is set to $p_k=-40 \text{ dBm},k =1,\cdots,K$. $\text{SNR}_{\text{n}} = {p_k}/{\sigma^2_n}$ denotes SNR at the SIM.

Furthermore, the number of grids for coarse estimation and fine estimation is set to $G_a = 32$, $G_e=16$, $V_a=16$, $V_e=8$, respectively. The number of SIM configurations used for coarse estimation and fine estimation is $Q=6$ and $Q=2$, respectively. For the EMNNs used for coarse estimation and fine estimation, the sizes of their training set $\{\hat{\bm\theta}\}$ are 50,000 and 100,000, respectively. The former is trained for 250 epochs while the latter is trained for 150 epochs. The batch size for both is set to 128. The perturbation SNR, i.e., $\text{SNR}_{\text{p}}={p_{\text{trn}}}/{\sigma^2_p}$, during training is set to $\text{SNR}_{\text{p}}=10$ dB. All EMNNs are trained by using the PyTorch framework and the Adam optimizer \cite{kingma2014adam}. The learning rate and the two parameters related to momentum are respectively set to $l_r = 0.001$, $\beta_1 = 0.9$, and $\beta_2 = 0.999$\cite{kingma2014adam}.

Next, we will introduce three metrics used to evaluate the performance of the DOA estimator in this paper, including classification error, RMSE, and peak-to-average difference of angular spectrum.

\begin{figure*}[!t]
	\centering
	\subfigure[]{
			\includegraphics[width=0.23\textwidth]{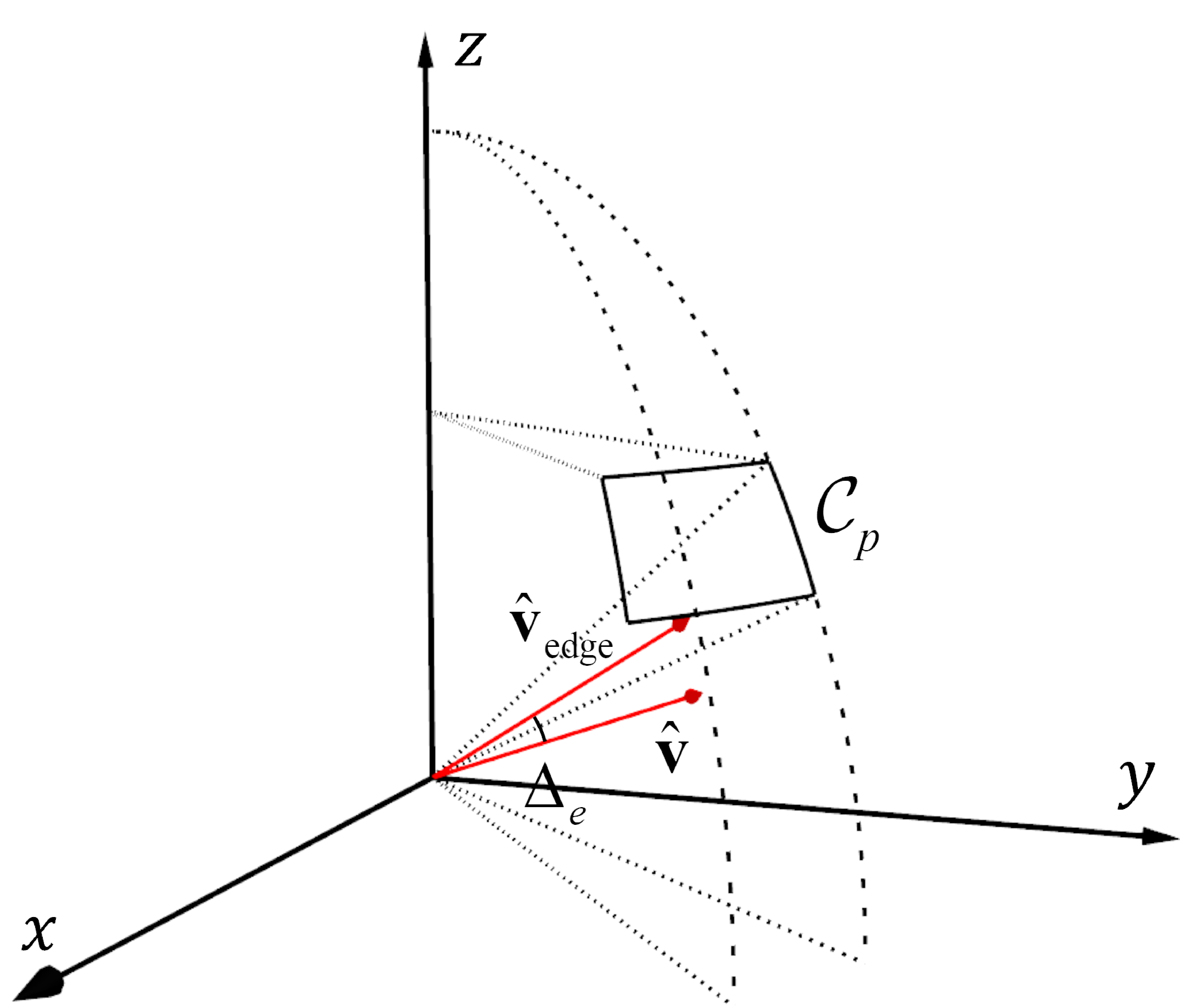}}
	\subfigure[]{
			\includegraphics[width=0.23\textwidth]{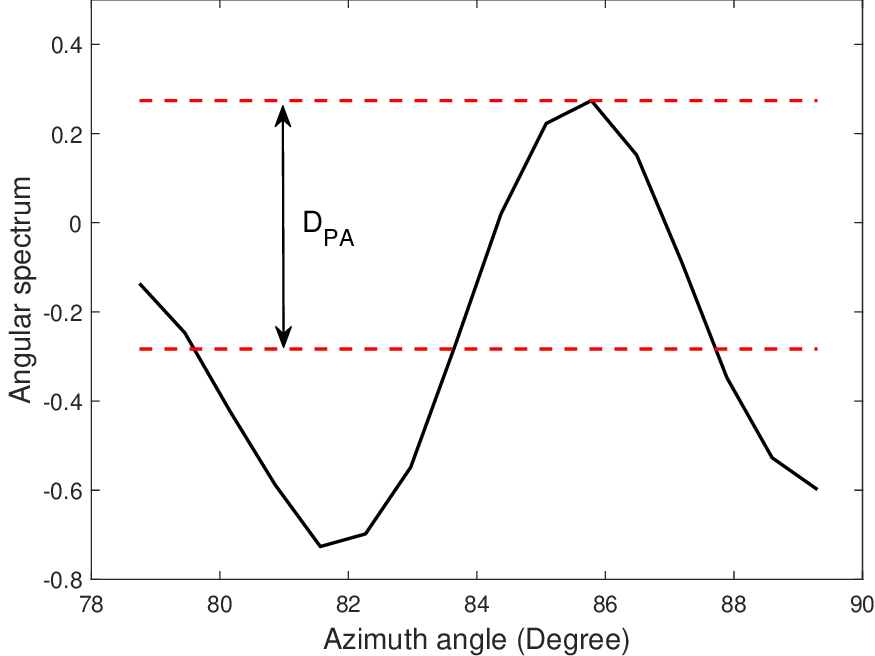}}
	\subfigure[]{
			\includegraphics[width=0.23\textwidth]{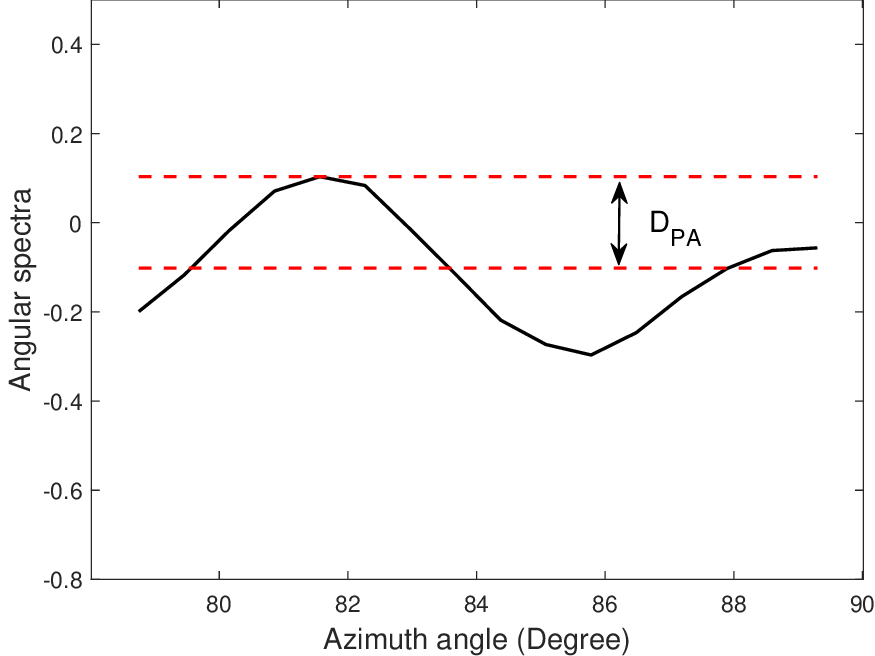}}
		\subfigure[]{
		\includegraphics[width=0.23\textwidth]{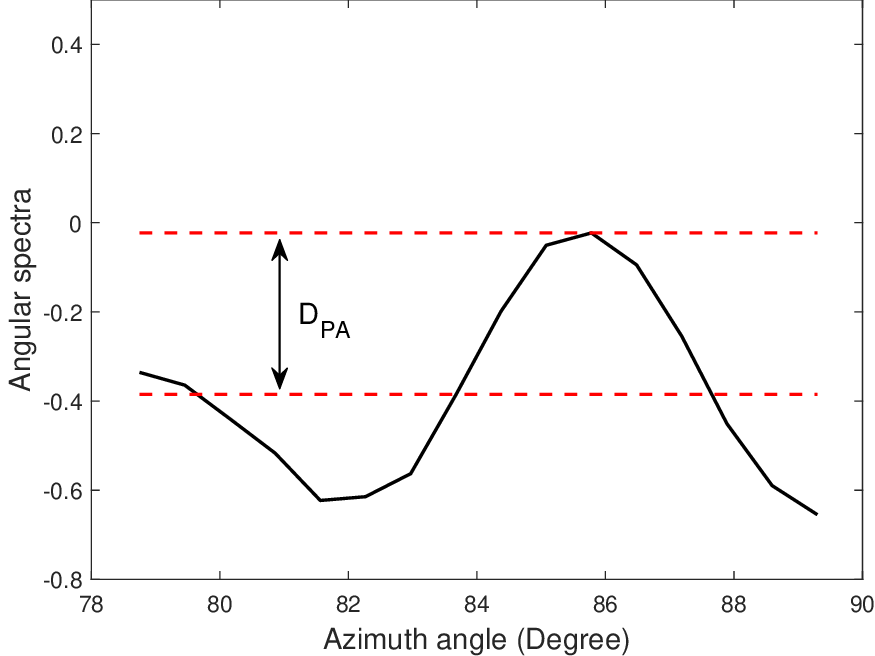}}
	\caption{(a) Illustration of classification error. (b) Angular spectrum generated by a signal incident from the estimated subregion, where $D_{\text{PA}}$ of the spectrum should be as large as possible. (c) Angular spectrum generated by a signal outside the estimated subregion, where $D_{\text{PA}}$ of the spectrum should be as small as possible. (d) Angular spectrum of two incident signals by superimposing (b) and (c), where $D_{\text{PA}}$ of the spectrum should be as large as possible.}
	\label{fig_5}\vspace{-0.2cm}
	\end{figure*}
\begin{figure*}[!t]
\centering
\subfigure[]{
		\includegraphics[width=0.32\textwidth]{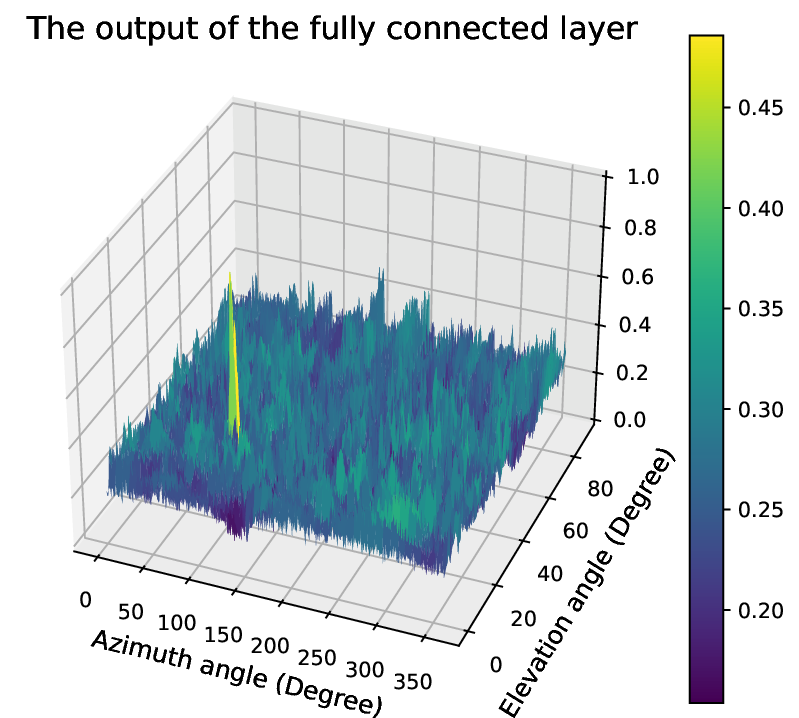}}
\subfigure[]{
		\includegraphics[width=0.32\textwidth]{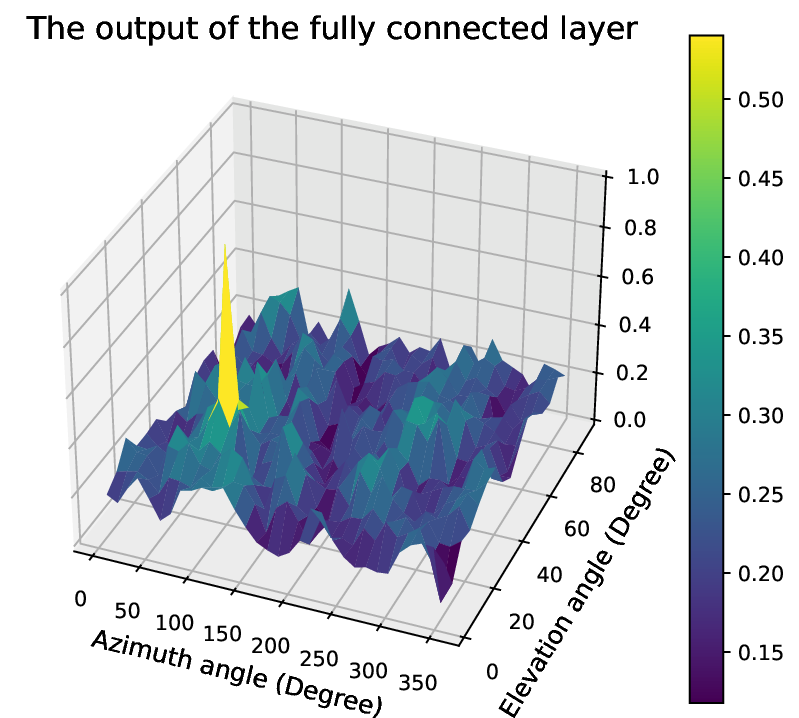}}
\subfigure[]{
		\includegraphics[width=0.32\textwidth]{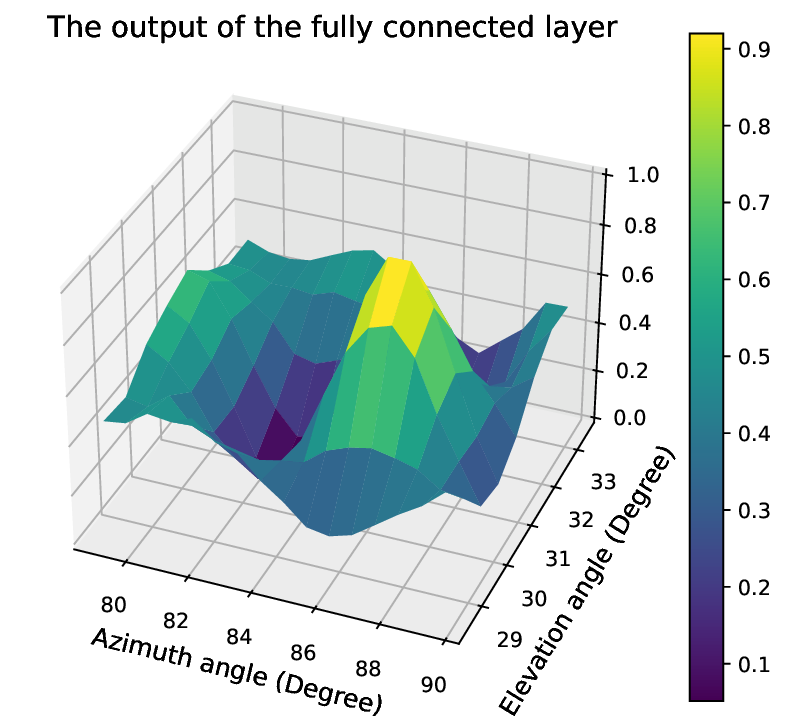}}
\caption{Generated spectra by EMNNs when $\theta^a=86.5^\circ, \theta^e=30.5^\circ$, (a) High-resolution global spectrum; (b) Coarse-resolution spectrum in the coarse estimation stage; (c) Fine-resolution spectrum in the fine estimation stage. }
\label{fig_6}\vspace{-0.5cm}
\end{figure*}

\subsubsection {Classification Error}
As shown in Fig. \ref{fig_5}(a), we first define an intermediate variable $\hat{\bm\theta}_\text{edge}$ to represent the DOA within the classified subregion $\mathcal{C}_p$ having the closest distance to the testing sample $\hat{\bm\theta}$, and use the angle between the directional vectors of these two DOAs to characterize the classification error. Specifically, $\hat{\bm\theta}_\text{edge}$ is represented as
\begin{equation}
\label{edge_model}
\hat{\bm\theta}_\text{edge}=\min \limits_{{\bm\theta}}\{||{\bm\theta}-\hat{\bm\theta}||_F^2,{\bm\theta}\in\mathcal{C}_p\}.
\end{equation}

Therefore, the classification error is defined as
\begin{equation} 
\Delta_{e}= \cos^{-1}( {\hat {\bf v}} \cdot \hat {\bf v}_\text{edge} ) \in [0,\pi],
\end{equation}
where ${\bf \hat v} = (\cos \hat{\theta}^a\sin \hat{\theta}^e,\sin \hat{\theta}^a\sin \hat{\theta}^e, \cos \hat{\theta}^e)^T$ denotes the direction vector corresponding to $\hat{\bm\theta}$ and ${\bf \hat v_\text{edge}} = (\cos \hat{\theta}^a_\text{edge}\sin \hat{\theta}^e_\text{edge},\sin \hat{\theta}^a_\text{edge}\sin \hat{\theta}^e_\text{edge}, \cos \hat{\theta}^e_\text{edge})^T$ denotes the direction vector corresponding to $\hat{\bm\theta}_\text{edge}$. When the classification result is correct, we have $\Delta_{e}=0^{\circ}$. Conversely, a larger classification error indicates that the testing sample is farther from the classified subregion.

Next, we illustrate the expression of \eqref{edge_model} by utilizing the coarse-resolution subregions in the coarse estimation. Suppose that the test sample $\hat{\bm\theta}$ is classified into subregion $\mathcal{G}_g$, $\hat{\bm\theta}_\text{edge}$ can be given by 
\begin{subequations}
\begin{align}
\hat{\theta}_\text{edge}^e&=\begin{cases}g^eE^{e,\text{coarse}}, &\text{if}\ \hat{\theta}^e \in (g^eE^{e,\text{coarse}},\pi],\\ \hat{\theta}^e,&\text{if}\ \hat{\theta}^e\in[\theta_{g}^{e,\text{start}},g^eE^{e,\text{coarse}}],\\ \theta_{g}^{e,\text{start}}, &\text{if}\ \hat{\theta}^e \in[0,\theta_{g}^{e,\text{start}}), \end{cases}\\
\hat{\theta}_\text{edge}^a&=\begin{cases}g^aE^{a,\text{coarse}}, &\text{if}\ \hat{\theta}^a \in \mathcal{D}_g^{\text{right}},\\ \hat{\theta}^a,&\text{if}\ \hat{\theta}^a\in[\theta_{a,\text{start}}^{g},g^aE^{a,\text{coarse}}],\\ \theta_{g}^{a,\text{start}}, &\text{if}\ \hat{\theta}^a \in\mathcal{D}_g^{\text{left}}, \end{cases}
\end{align}
\end{subequations}
where $\mathcal{D}_g^{\text{right}}=(g^aE^{a,\text{coarse}},\theta^{a,\text{start}}_g+E^{a,\text{coarse}}/2+\pi]$, $\mathcal{D}_g^{\text{left}}=[0,\theta_{g}^{a,\text{start}})\cup(\theta^{a,\text{start}}_g+E^{a,\text{coarse}}/2+\pi,2\pi]$ for $\theta^{a,\text{start}}_g+E^{a,\text{coarse}}/2 < \pi$ and $\mathcal{D}_g^{\text{right}}=[0,\theta^{a,\text{start}}_g+E^{a,\text{coarse}}/2-\pi)\cup (g^aE^{a,\text{coarse}},2\pi]$, $\mathcal{D}_g^{\text{left}}=[\theta^{a,\text{start}}_g+E^{a,\text{coarse}}/2-\pi,\theta^{a,\text{start}}_g)$ for $\theta^{a,\text{start}}_g+E^{a,\text{coarse}}/2 > \pi$.

\subsubsection{RMSE}
The RMSE is defined as
\begin{equation}\label{fixed_angle_rmse_define_v3}
\Delta_{\rm E}
=\sqrt{(\Delta\theta^a)^2+(\Delta\theta^e)^2},
\end{equation}
where
\begin{subequations}
\begin{align}
\Delta\theta^a
&=\min\!\left\{|\hat\theta^a-\theta^a|,\,2\pi-|\hat\theta^a-\theta^a|\right\},\label{delta_a_def}\\
\Delta\theta^e&=\hat\theta^e-\theta^e.\label{delta_e_def}
\end{align}
\end{subequations}

\subsubsection {Peak-to-Average Difference of Angular Spectrum}
We employ the peak-to-average difference to quantify the prominence of the peak in the angular spectrum, which is denoted as
\begin{equation}
\label{DPM_define}
D_\text{PA}=\max\{{\bf f}\}-\frac{\sum_{p=1}^P[{\bf f}]_p}{P}.
\end{equation}
In the fine estimation phase, $D_\text{PA}$ holds two different implications for the testing sample $\hat{\bm\theta}$. 
\begin{itemize}
\item If $\hat{\bm\theta}\in\mathcal{G}_g$, a larger value of $D_{\text{PA}}$ is expected when the peak location of the sub-spectrum aligns with $\hat{\bm\theta}$, this implies that other positions in the angular spectrum hardly surpass the spectral peak, as shown in Fig. \ref{fig_5}(b).
\item If $\tilde{\bm\theta}\notin\mathcal{G}_g$, as shown in Fig. \ref{fig_5}(c)-(d), the signal component outside the subregion $\mathcal{G}_g$ would affect the estimation of DOA within $\mathcal{G}_g$. Specifically, greater spectral fluctuations indicate stronger interference. 
\end{itemize}
Hence, as a result, a higher value of $D_{\text{PA}}$ is expected when the signal DOA is aligned with $\hat{\bm\theta}$, and the value of $D_{\text{PA}}$ should be reduced when the signal is from the outside subregion $\mathcal{G}_g$.

\subsection{Simulation Results}
Fig. \ref{fig_6} illustrates the angular spectra generated by the EMNN when considering $\theta^a=86.5^{\circ},\theta^e=30.5^{\circ}$. Specifically, the high-resolution spectrum in Fig. \ref{fig_6}(a) is obtained by considering grid division of $G_a=512, G_e=128$, while Fig. \ref{fig_6}(b) and Fig. \ref{fig_6}(c) show the spectra corresponding to the two stages of the proposed hierarchical estimation protocol. It can be observed that the spectra exhibit a prominent peak in the region corresponding to the incident angle. Compared to directly using the EMNN for generating the global high-resolution spectrum, the proposed hierarchical estimation protocol is capable of achieving the same classification resolution, while significantly reducing the computational complexity.

\begin{figure}[!t]
\centering
\subfigure[$\theta^a=30^\circ, \theta^e=60^\circ$]{
		\includegraphics[width=0.23\textwidth]{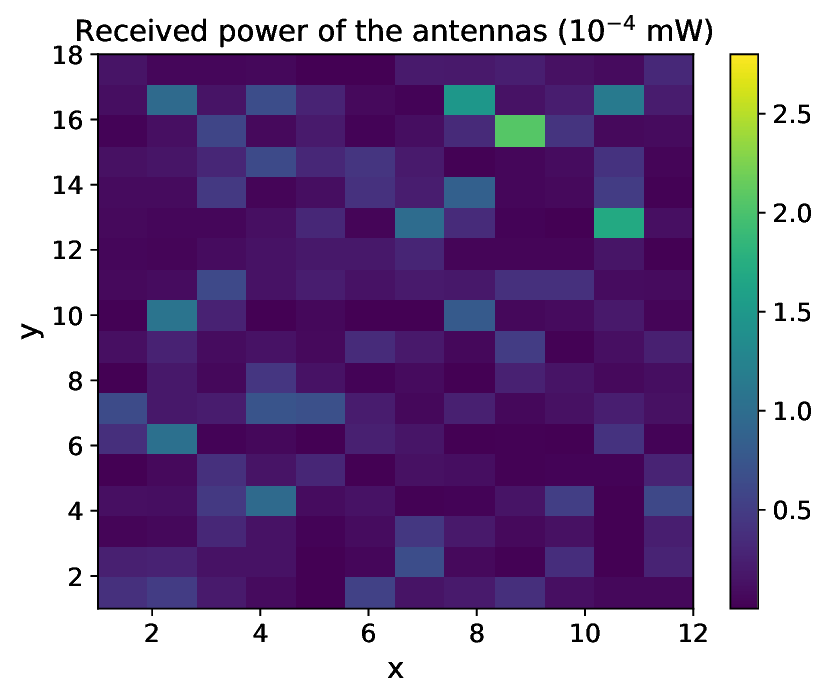}}
\subfigure[$\theta^a=0^\circ, \theta^e=45^\circ$]{
		\includegraphics[width=0.23\textwidth]{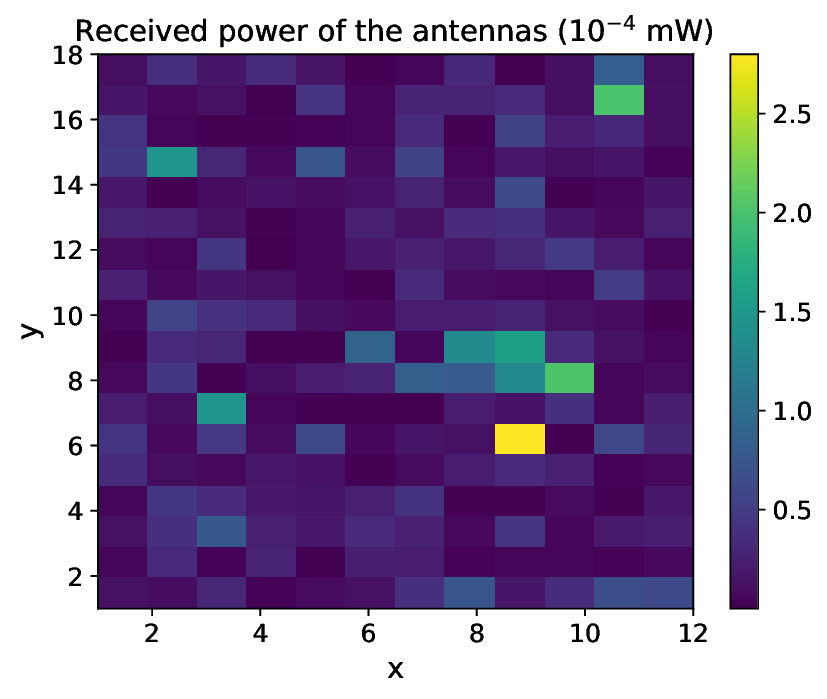}}
\caption{Received power pattern across antenna array. }
\label{fig_7}\vspace{-0.2cm}
\end{figure}

Fig. \ref{fig_7}(a) and Fig. \ref{fig_7}(b) illustrate the distribution of received power across the antenna array when the incident angles $(\theta^a,\theta^e)$ are $(30^{\circ},60^{\circ})$ and $(0^{\circ},45^{\circ})$, respectively. The power distribution of UPA formed by the six SIM phase shift configurations is arranged in three rows and two columns. It can be seen that the received energy distributions exhibit clear clustering phenomena. This implies that the SIM extracts the amplitude-phase distribution characteristics of the incident signals and transforms them into the received power distribution, thereby enabling DOA estimation based solely on amplitude measurements. Conversely, in an array system without the SIM, the received power distribution is constant for a single incident signal, and the DOA can not be estimated by leveraging low-complexity amplitude detection.

\begin{figure}[!t]
\centering
\subfigure[]{
		\includegraphics[width=0.23\textwidth]{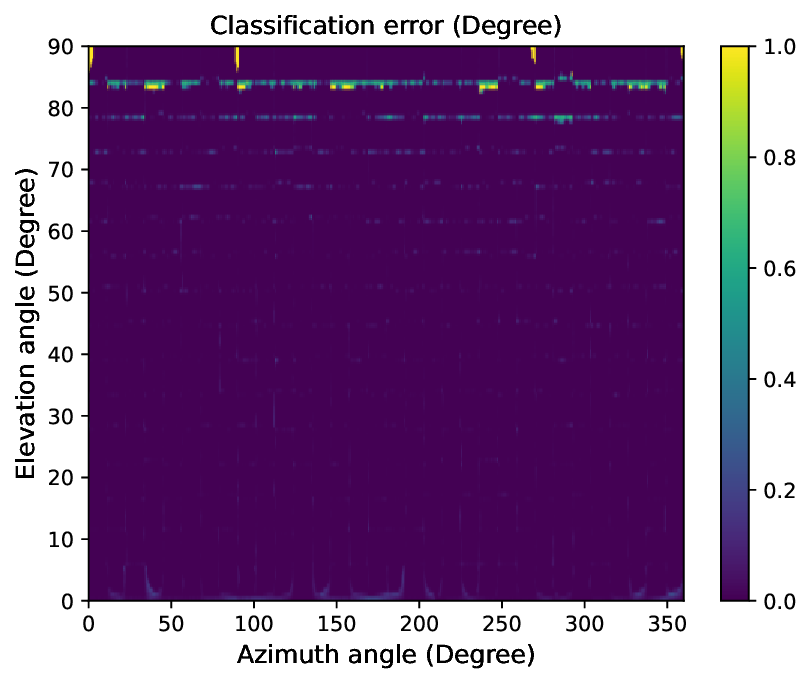}}
\subfigure[]{
		\includegraphics[width=0.23\textwidth]{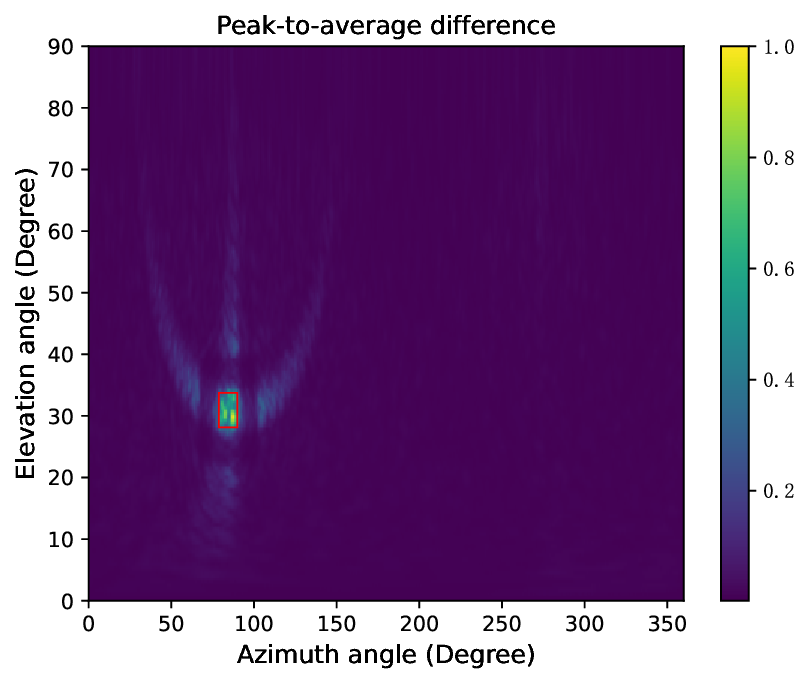}}
\caption{(a) The classification error versus DOA for coarse estimation. (b) The peak-to-average difference $D_\text{PA}$ versus DOA for fine estimation. }
\label{fig_8}\vspace{-0.5cm}
\end{figure}

In Fig. \ref{fig_8}(a), we evaluate the classification error versus DOA for coarse estimation under SNR$_\text{n}=5$ dB and $S$ = 10. Specifically, the entire DOA region is segmented into $512\times128$ pixels for azimuth and elevation angles. Within each pixel, 100 samples are randomly drawn to compute the mean classification error of misclassified samples for visualization. As seen in Fig. \ref{fig_8}(a), the EMNN accurately classifies DOAs of incident signals in most cases, with $96\%$ samples perfectly classified into the corresponding DOA subregions. Furthermore, the estimation error becomes significant in high-elevation-angle regions. Nonetheless, the DOAs of targets in these regions are primarily misclassified into adjacent subregions, thus preventing significant errors. In Fig. \ref{fig_8}(b), we evaluate the peak-to-average difference $D_{\text{PA}}$ of the sub-spectrum generated by a specific EMNN for fine estimation. Specifically, the EMNN is trained for the subregion ${\mathcal{G}}_{ g}$, with $g^a=8, g^e=6$, corresponding to the red box in Fig. \ref{fig_8}(b). After inputting the signal from each potential direction into the EMNN, the peak-to-average difference of the output sub-spectrum is calculated and normalized. Notably, the magnitude of $D_\text{PA}$ within the ${\mathcal{G}}_{ g}$ is significantly higher than that of signals from other regions, highlighting the efficacy of the proposed labeling approach in mitigating interference from signals outside the current subregion. 


\begin{figure}[!t]
	\centering
	\includegraphics[width=0.4\textwidth]{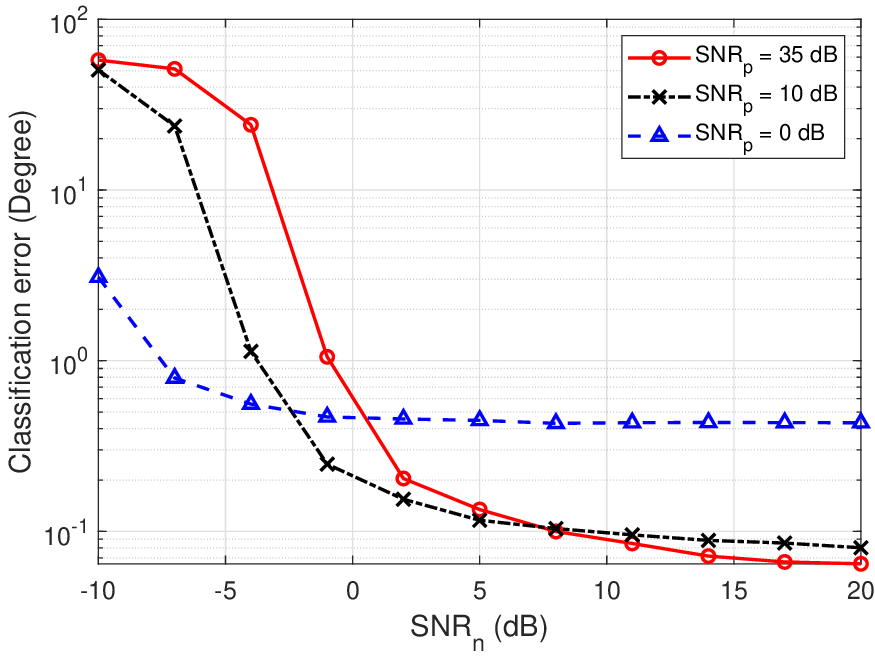}
	\caption{Classification error for coarse estimation versus $\text{SNR}_{\text{n}}$.}
	\label{fig_9}\vspace{-0.2cm}
\end{figure}

In Fig. \ref{fig_9}, we illustrate the classification error for coarse estimation versus $\text{SNR}_{\text{n}}$ for $G_a=32$, $G_e=16$, $L=5$, $N=1600$, $Q=3$. Moreover, we consider varying $\text{SNR}_{\text{p}}$ values during the training stage. Note that classification error improves as $\text{SNR}_{\text{n}}$ increases. Although the introduction of perturbation noise during training enhances robustness against measurement errors, it is suggested to accommodate varying SNR levels encountered in practical scenarios. For example, although low perturbation SNR during training leads to higher DOA estimation accuracy in low SNR environments, it diminishes accuracy in high SNR environments.

\begin{figure}[!t]
\centering
\includegraphics[width=0.4\textwidth]{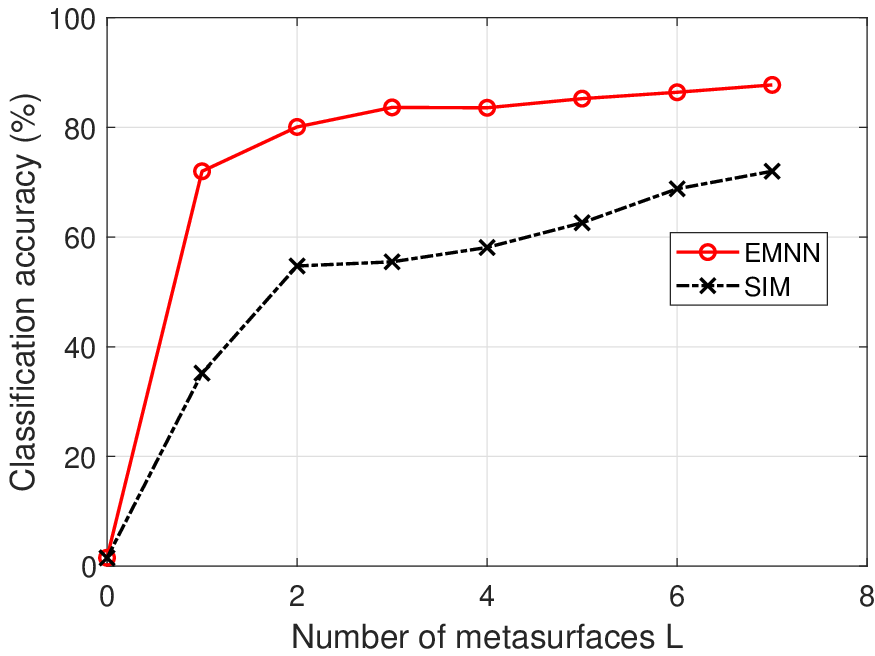}
\caption{Classification accuracy for coarse estimation versus the number of metasurface layers $L$.}
\label{fig_10}\vspace{-0.5cm}
\end{figure}

In Fig. \ref{fig_10}, we present the classification accuracy of coarse estimation versus the number of layers $L$ in the single-signal scenario, where we set $\text{SNR}_{\text{n}}=35$ dB, $\text{SNR}_{\text{p}}=35$ dB, $G_a=8$, $G_e=8$, $N=225$, $M=16$, $S = 1$, $Q = 1$. Meanwhile, we compare the DOA estimation performance of the EMNN to plain SIM. In the SIM-aided system, each receiving antenna is associated with a specific DOA region, and the DOA region corresponding to the receiving antenna with the highest received power is designated as the corresponding class. Therefore, the number of receiving antennas required for the SIM-aided system is $G_a\times G_e = 64$. Notably, to ensure that adjacent receiving antennas correspond to adjacent DOA subregions within the SIM system, and considering the periodicity of the azimuth angle, $64$ antennas are arranged in $8$ radially equidistant concentric circular arrays. For each array, the angular spacing between antennas is set as $2 \pi /G_a$, and the radial spacing between adjacent circular arrays is set as $\lambda / 2$. Specifically, when $L=0$, the received signal power is directly utilized for training and extracting the peak for DOA estimation. From Fig. \ref{fig_10}, it can be observed that as the number of metasurface layers increases, both the classification accuracy of the EMNN and the SIM improve due to the enhanced inference capability of the SIM. When $L=0$, the uniform power distribution is observed, thus yielding random classification results for both two schemes. Furthermore, the proposed EMNN consistently outperforms the SIM in terms of classification accuracy, indicating the gain achieved through the integration of the fully connected layer. 


\begin{figure}[!t]
\centering
\subfigure[]{
		\includegraphics[width=0.23\textwidth]{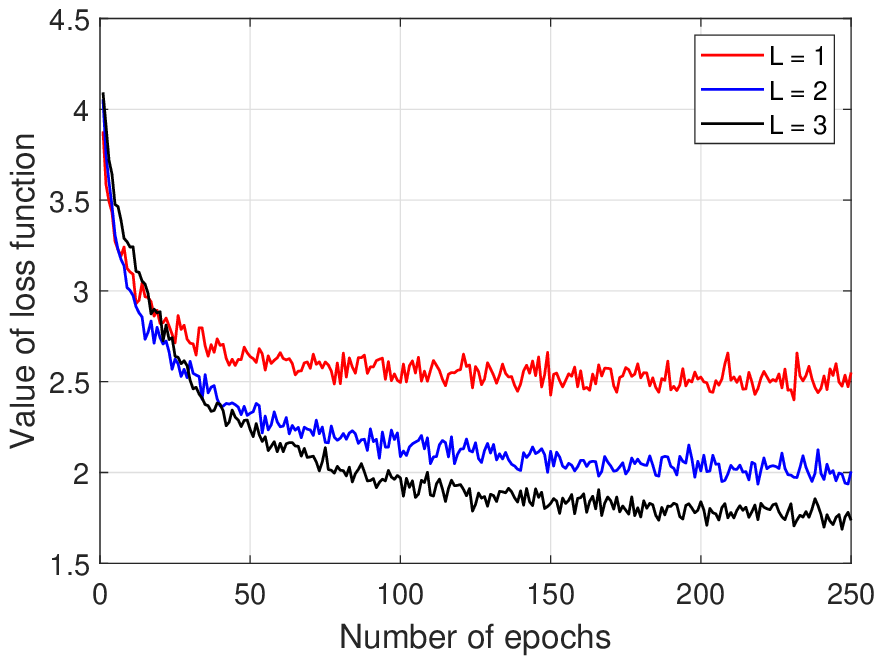}}
\subfigure[]{
		\includegraphics[width=0.23\textwidth]{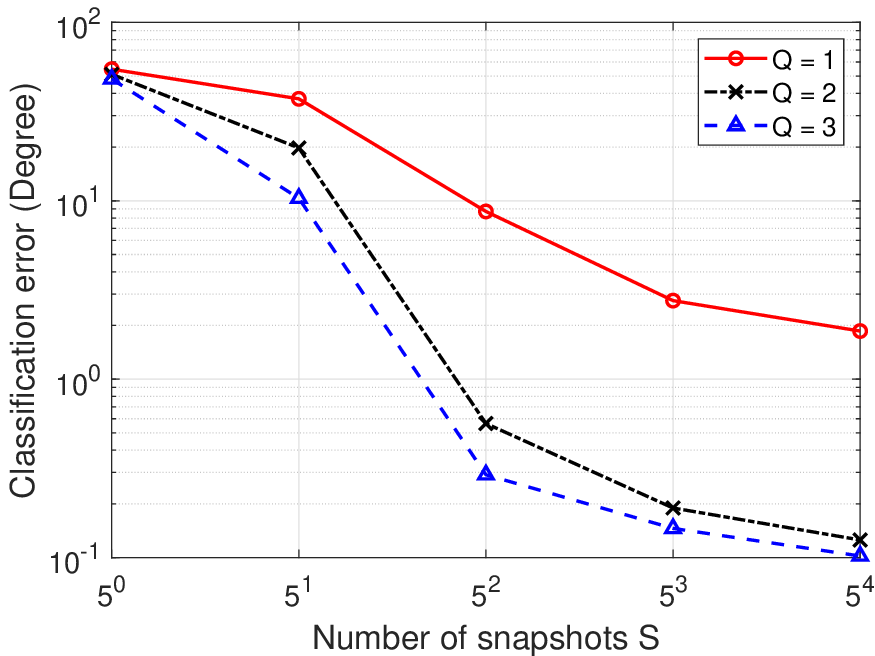}}
\caption{(a) The loss function in the training process. (b) Classification error for coarse estimation versus the number of snapshots $S$. }
\label{fig_11}\vspace{-0.2cm}
\end{figure}

Additionally, in Fig. \ref{fig_11}(a), we examine the convergence of the loss function during the training process of the EMNN for various $L$. It is noted that the loss function converges by increasing the number of training epochs under all configurations. In Fig. \ref{fig_11}(b), we present the classification error of coarse estimation versus the number of snapshots $S$ in single-signal scenario for $\text{SNR}_{\text{n}}=0$ dB, $\text{SNR}_{\text{p}}=10$ dB, $G_a=32$, $G_e=16$, $L=5$, $N=1600$, $M=16$, where the varying numbers of SIM configurations $Q$ is considered. It can be observed that the classification error decreases as the number of snapshots increases, thanks to the noise smoothing effect. Moreover, the EMNN utilizing multiple configurations exhibits stronger DOA estimation capability compared to those employing the single SIM configuration. This indicates the advantage of employing multiple SIM configurations in EMNNs.

\begin{figure}[!t]
\centering
\includegraphics[width=0.4\textwidth]{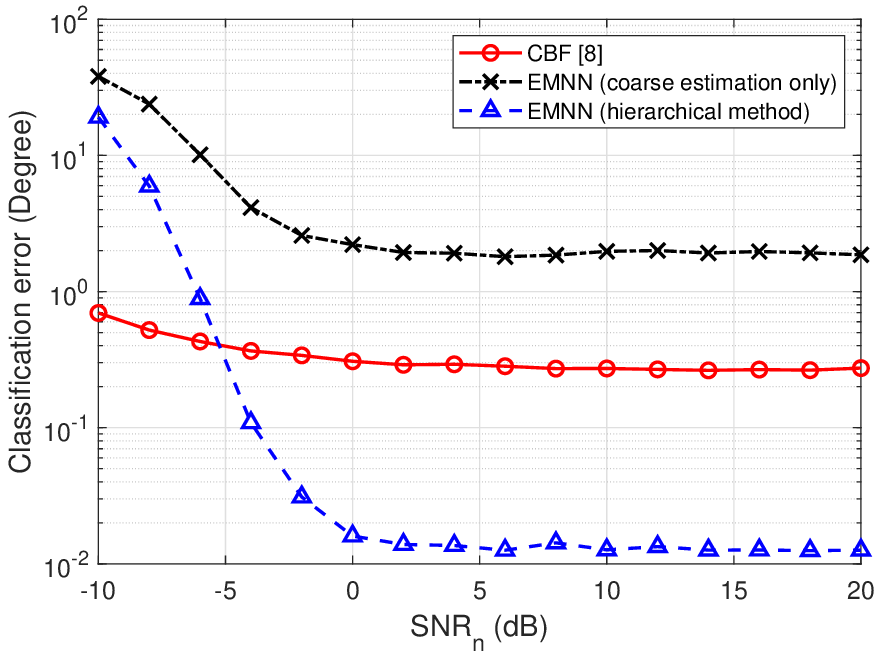}
\caption{Classification error versus $\text{SNR}_{\text{n}}$ in dual-signals scenario.}
\label{fig_12}\vspace{-0.5cm}
\end{figure}

Fig. \ref{fig_12} illustrates the classification error in a dual-signal scenario for $S=100$, where we contrast the proposed method with the coarse estimation only and the CBF method. For the sake of comparison, we use the same grid resolution for the coarse estimation only and the hierarchical method, i.e., $G_a=512$, $G_e=128$. Moreover, the CBF method utilizes an identical antenna array as the EMNN system. Additionally, to avoid ambiguity, the antenna spacing is set to $d_x^{\text{A}}=d_y^{\text{A}}=\lambda/2$ for the CBF method. For each signal sample, the two testing DOAs are randomly chosen from ${\mathcal{G}}_{ g}, g^a=8, g^e=6$ and ${\mathcal{G}}_{ \hat g},\hat{g}^a=12,\hat{ g}^e=6$, respectively. Moreover, the classification error is obtained by computing the average errors associated with two pairs between the estimated results and the testing DOAs and selecting the smaller value. From Fig. \ref{fig_12}, it can be observed that the hierarchical method exhibits significantly smaller classification error when compared to using coarse estimation only. This indicates that the introduction of the fine estimation stage effectively improves the DOA estimation accuracy. When comparing with the CBF method, it is observed that, under lower SNR, the performance of the EMNN-based method is inferior to the CBF method, and under higher SNR, the EMNN-based method outperforms the CBF method. This discrepancy arises because, in the low SNR region, the classification errors primarily result from peak detection being affected by noise. Due to energy attenuation caused by multi-layer propagation, the EMNN method performs worse than the CBF method. As the SNR increases, classification errors are predominantly determined by signal interference rather than noise. Since the EMNN demonstrates a stronger capability in mitigating signal interference due to the higher angular resolution provided by the large aperture of the metasurface, its performance significantly surpasses that of the CBF method under high-SNR conditions.

\begin{figure}[!t]
\centering
\includegraphics[width=0.4\textwidth]{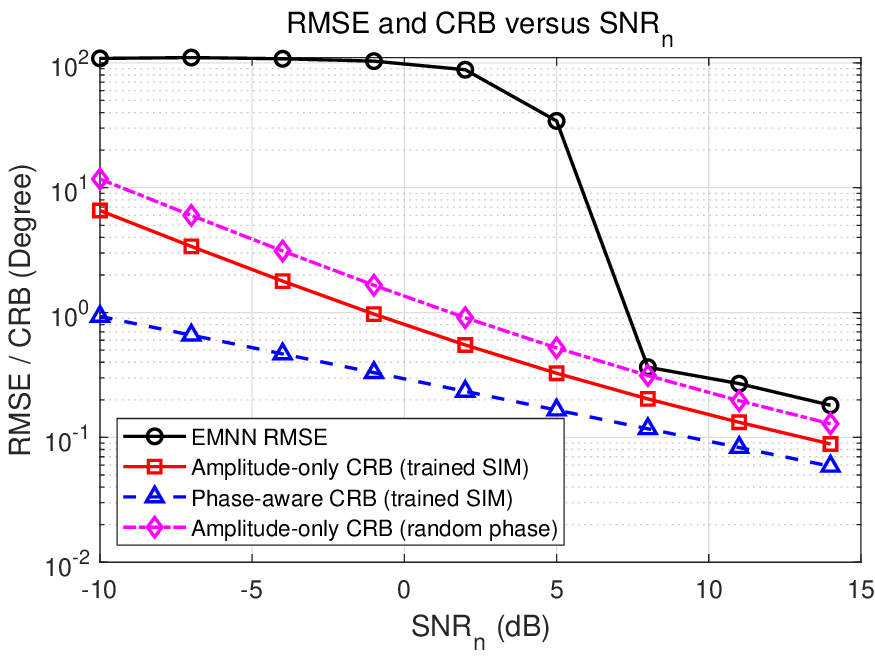}
\caption{RMSE and CRB versus $\text{SNR}_{\text{n}}$ at the fixed DOA $\theta^a=30.1^{\circ}, \theta^e=44.8^{\circ}$.}
\label{fig_13}\vspace{-0.5cm}
\end{figure}

In Fig. \ref{fig_13}, we evaluate the RMSE of the coarse estimation versus $\text{SNR}_{\text{n}}$ at the fixed testing DOA $\theta^a=30.1^{\circ}, \theta^e=44.8^{\circ}$, alongside the amplitude-only CRB with the trained SIM, the amplitude-only CRB under random SIM phase configurations, and the phase-aware CRB. The simulation parameters are set as $G_a=1024$, $G_e=256$, $Q=9$, $S=1$, where the EMNN is trained on the single-signal dataset, and the DOA estimate is taken as the center of the predicted grid. The integrals involved in evaluating the CRB are computed via the numerical integration routines from the SciPy library. It can be observed that the RMSE exceeds the amplitude-only CRB with the trained SIM, as a single fully connected layer is insufficient to attain the bound. Nonetheless, the trained SIM yields a lower amplitude-only CRB than random configurations, indicating its effectiveness in preserving the angle-relevant information when converting the wavefront into the received power distribution. The phase-aware CRB attains the lowest value among the three cases, consistent with the fact that retaining the phase incurs less information loss.

\section{Conclusions}
In this paper, we propose a DOA estimation method by leveraging the advanced EMNN, which consists of a UAV-mounted SIM and a fully connected layer to generate the angular spectrum. Furthermore, to reduce the computational burden and the requirement for observation snapshots, a hierarchical DOA estimation method was proposed. Moreover, we derived the single-snapshot CRB of the SIM-based sensing receiver under amplitude-only observation. Simulation experiments validate the effectiveness of the proposed EMNN-based DOA estimators and demonstrate that, under certain conditions, the proposed method outperforms the CBF algorithm with the same computational complexity. In summary, leveraging the powerful wave-domain computing capabilities of SIM, the proposed DOA estimation method not only reduces hardware complexity but also maintains high performance with lower digital computing requirements, showing potential applications in UAV communication scenarios where digital computing capability is limited.

Furthermore, leveraging the contributions of this paper, the proposed EMNN-based DOA estimation methodology can be advanced in several areas for future research, including: i) Employing techniques such as transfer learning to enable the EMNN to rapidly adapt to more challenging scenarios, including those with coherent signals or hardware impairments. ii) Integrating non-linear components, for instance, power amplifiers\cite{NE_2023_Gao_Programmable}, within the meta-atoms to further augment the inference capabilities of the EMNN. iii) Introducing unsupervised learning and self-feedback mechanisms to refine the network architecture and improve estimation performance. Moreover, extending the SIM concept to low-latency localization at the EM domain is an intriguing avenue for future work. This could involve leveraging time-varying characteristics of SIMs, for instance, by designing real-time waveforms to perform distance or velocity estimation within the EM domain.

However, numerous challenges remain. From a training methodology perspective, future work needs to address the limitations of current models that assume continuous phase modulation and perfectly known inter-layer propagation coefficients. This necessitates exploring optimization algorithms for discrete variables and designing error calibration algorithms.

\appendices
\section{CRB Under Amplitude-Only Observation}
For the per-antenna observation $y_{m,q}=\alpha_{m,q}({\bm\theta})s+n_{m,q}$ extracted from the signal model in \eqref{y_model}, $2z_{m,q}$ follows a non-central chi-square distribution with two degrees of freedom and non-centrality parameter $2\gamma_{m,q}({\bm\theta})$, whose probability density function reads \cite{nistdlmf,johnson1995continuous2}
\begin{equation}\label{power_pdf_appendix_v3}
f_{Z}(z;\gamma)=e^{-(z+\gamma)}I_{0}(2\sqrt{\gamma z}),\ z\ge 0.
\end{equation}

Accordingly, the score function with respect to $\gamma$ is
\begin{equation}\label{score_appendix_v3}
\frac{\partial\ln f_{Z}(z;\gamma)}{\partial\gamma}
=\sqrt{z/\gamma}\,\frac{I_{1}(2\sqrt{\gamma z})}{I_{0}(2\sqrt{\gamma z})}-1,
\end{equation}
and the per-sample scalar Fisher information is
\begin{equation}\label{j1_appendix_v3}
{\mathcal J}_{1}(\gamma)
={\mathbb E}_{Z}\!\left[\left(\frac{\partial\ln f_{Z}(z;\gamma)}{\partial\gamma}\right)^{2}\right],
\end{equation}
which, upon substituting \eqref{power_pdf_appendix_v3} and \eqref{score_appendix_v3}, coincides with \eqref{J1_def}. Differentiating $\gamma_{m,q}({\bm\theta})$ and invoking \eqref{scalars} give
\begin{subequations}\label{app_lambda_grad}
\begin{align}
\partial\gamma_{m,q}/\partial\theta^{a}
&=2\rho\operatorname{real}\{\alpha_{m,q}^{*}\beta^{a}_{m,q}\}
=[{\bm\eta}_{m,q}({\bm\theta})]_{1},\\
\partial\gamma_{m,q}/\partial\theta^{e}
&=2\rho\operatorname{real}\{\alpha_{m,q}^{*}\beta^{e}_{m,q}\}
=[{\bm\eta}_{m,q}({\bm\theta})]_{2},
\end{align}
\end{subequations}
which identifies
${\bm\eta}_{m,q}({\bm\theta})=\nabla_{\bm\theta}\gamma_{m,q}({\bm\theta})$.

Since $\gamma_{m,q}({\bm\theta})$ enters \eqref{power_pdf_appendix_v3} only through the non-centrality parameter, the chain rule yields
\begin{equation}\label{app_chain_score}
\nabla_{\bm\theta}\ln f_{Z}(z_{m,q};\gamma_{m,q})
=\frac{\partial\ln f_{Z}(z;\gamma)}{\partial\gamma}\bigg|_{\substack{z=z_{m,q}\\ \gamma=\gamma_{m,q}}}\!{\bm\eta}_{m,q}({\bm\theta}).
\end{equation}

Taking the expected outer product of \eqref{app_chain_score} and invoking \eqref{j1_appendix_v3} reduce the per-sample FIM to the rank-one form
\begin{equation}\label{app_per_sample_fim}
{\mathbb E}_{Z}\!\left\{\nabla_{\bm\theta}\ln f_{Z}\,\nabla_{\bm\theta}^{T}\ln f_{Z}\right\}
={\mathcal J}_{1}(\gamma_{m,q}){\bm\eta}_{m,q}{\bm\eta}_{m,q}^{T}.
\end{equation}

Since $\{z_{m,q}\}_{m,q}$ are mutually independent, and the per-sample FIMs in \eqref{app_per_sample_fim} accumulate additively \cite{kay1993fundamentals}, giving
\begin{equation}\label{app_fpow_final}
{\bf F}_{\rm pow}({\bm\theta})
=\sum_{q=1}^{Q}\sum_{m=1}^{M}
 {\mathcal J}_{1}(\gamma_{m,q}({\bm\theta})){\bm\eta}_{m,q}({\bm\theta}){\bm\eta}_{m,q}^{T}({\bm\theta}),
\end{equation}
which recovers \eqref{crb_power_fim_remark_v3}.

\section*{Acknowledgment}
The authors would like to thank Jinbao Li for the helpful discussions.

\bibliographystyle{IEEEtran} 
\bibliography{ref}
\begin{IEEEbiography}[{\includegraphics[width=1in,height=1.25in,clip,keepaspectratio]{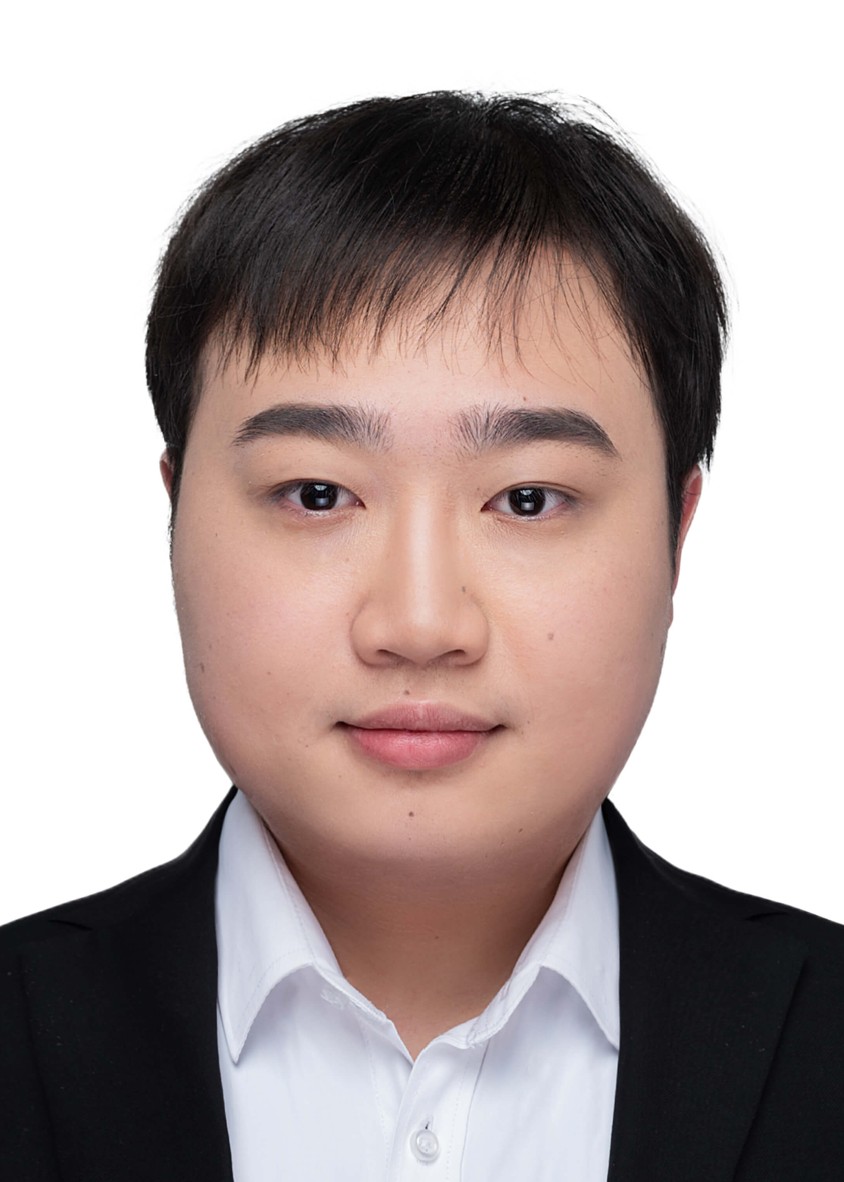}}]{Shining Lin}
received his B.S. degree in 2022 and M.S. degree in 2025, both from the School of Information and Communication Engineering at the University of Electronic Science and Technology of China (UESTC), Chengdu. His current research focuses on electromagnetic neural networks (EMNN), stacked intelligent metasurfaces (SIM), and DOA estimation. He was honored with the National Scholarship in 2024.
\end{IEEEbiography}

\begin{IEEEbiography}[{\includegraphics[width=1in,height=1.25in,clip,keepaspectratio]{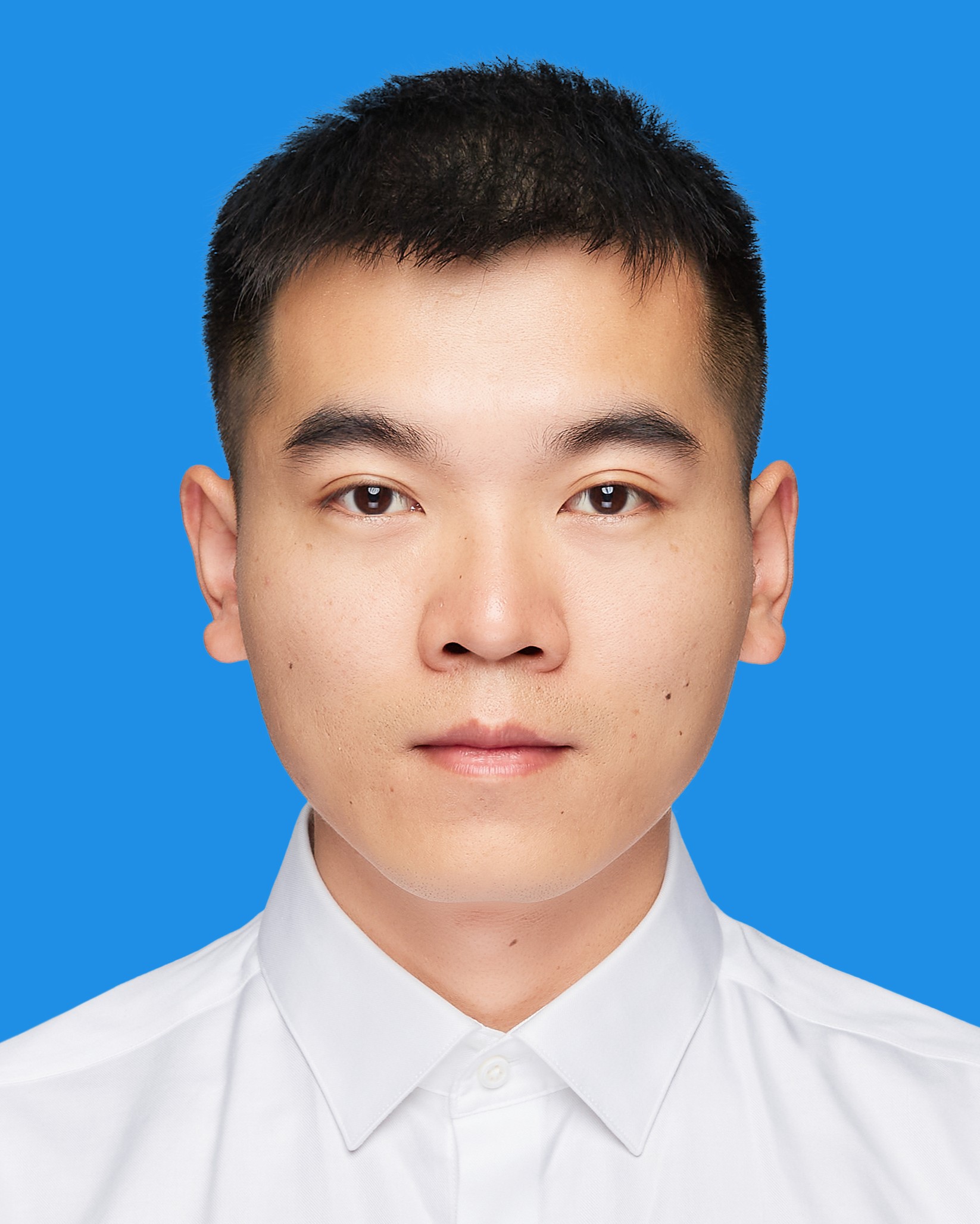}}]{Jiancheng An}
(Senior Member, IEEE) received his B.S. in Electronics and Information Engineering (2016) and Ph.D. in Information and Communication Engineering (2021) from the University of Electronic Science and Technology of China (UESTC), Chengdu. He was a Visiting Scholar at the University of Southampton, U.K. (2019–2020), a Post-Doctoral Fellow at Singapore University of Technology and Design (2021–2023), and a Research Fellow at Nanyang Technological University, Singapore (2023–2026). He is currently a Professor at UESTC's School of Electronic Science and Engineering. He received the IEEE ICC 2023 Best Paper Award, co-invented six patents, and has published over 100 papers in peer-reviewed journals and conferences. His research focuses on stacked intelligent metasurfaces (SIM), flexible intelligent metasurfaces (FIM), and electromagnetic neural networks (EMNN). He serves as an Editor for IEEE Transactions on Communications, IEEE Open Journal of the Communications Society, and IEEE Wireless Communications Letters, and as Lead Guest Editor for a Special Issue on SIM-empowered signal processing for 6G in IEEE Wireless Communications.
\end{IEEEbiography}

\begin{IEEEbiography}[{\includegraphics[width=1in,height=1.25in,clip,keepaspectratio]{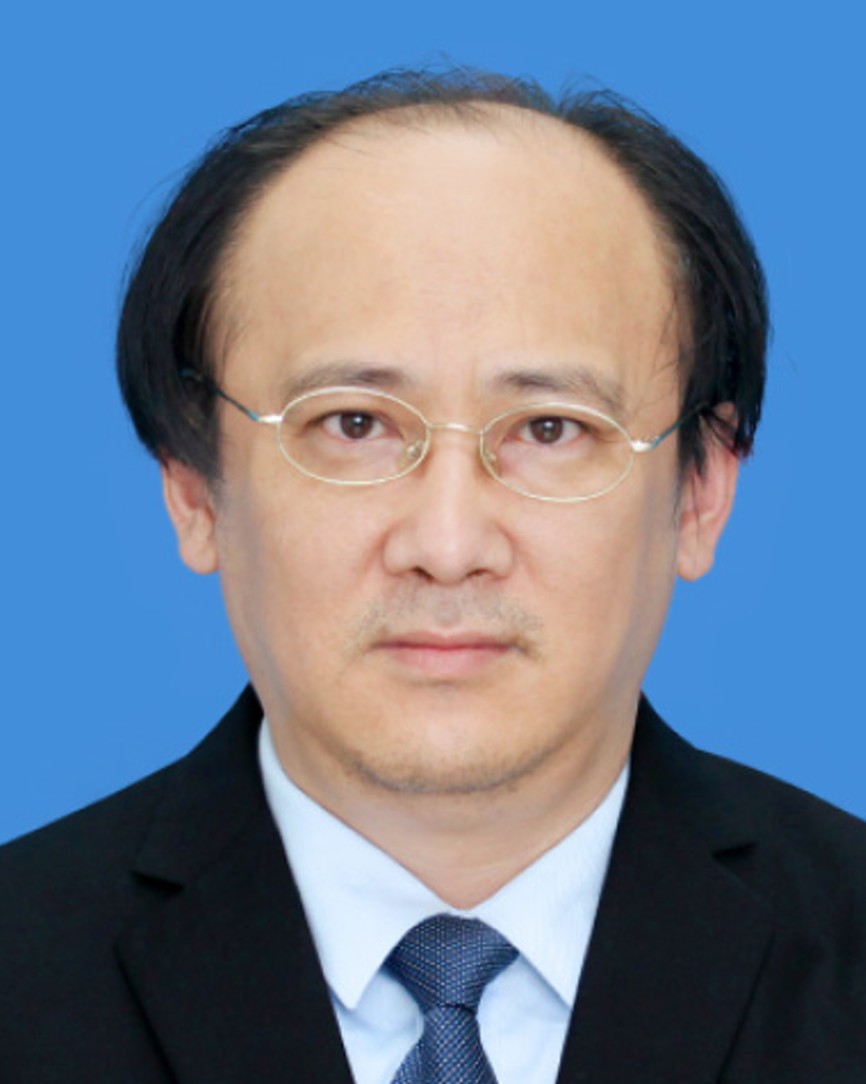}}]{Lu Gan}
(Member, IEEE) received the M.S. and Ph.D. degrees from the University of Electronic Science and Technology of China (UESTC), Chengdu, China, in 2002 and 2006, respectively. From September 2012 to September 2013, he was a Visiting Researcher with the University of Concordia, Montreal, QC, Canada. Since August 2014, he has been a Professor with UESTC. His research interests include signal detection and classification, array signal processing, compressive sensing, passive radar, and reconfigurable intelligent surfaces.
\end{IEEEbiography}

\begin{IEEEbiography}[{\includegraphics[width=1in,height=1.25in,clip,keepaspectratio]{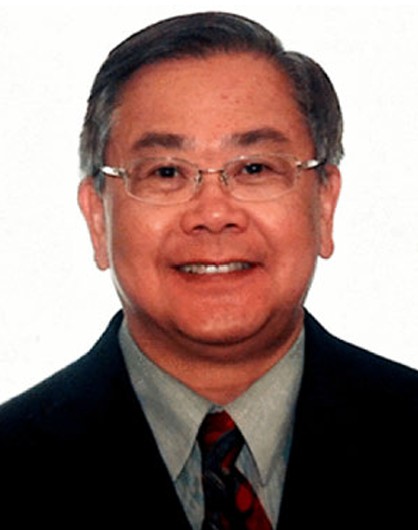}}]{Victor C. M. Leung}
(Life Fellow, IEEE) is a Distinguished Professor of computer science and software engineering with Shenzhen University, and also an Emeritus Professor of electrical and computer engineering and the Director of the Laboratory for Wireless Networks and Mobile Systems, The University of British Columbia (UBC), Canada. His published works have together attracted more than 80000 citations. His broad research interests include wireless networks and mobile systems. He has published widely in these areas. He is a fellow of the Royal Society of Canada (Academy of Science), Canadian Academy of Engineering, and the Engineering Institute of Canada. He received the 1977 APEBC Gold Medal, the 1977–1981 NSERC Postgraduate Scholarships, the IEEE Vancouver Section Centennial Award, the 2011 UBC Killam Research Prize, the 2017 Canadian Award for Telecommunications Research, the 2018 IEEE TCGCC Distinguished Technical Achievement Recognition Award, and the 2018 ACM MSWiM Reginald Fessenden Award. He co-authored papers that won the 2017 IEEE ComSoc Fred W. Ellersick Prize, the 2017 IEEE Systems Journal Best Paper Award, the 2018 IEEE CSIM Best Journal Paper Award, and the 2019 IEEE TCGCC Best Journal Paper Award. He was named in the Clarivate Analytics list of “Highly Cited Researchers” for several years. He is serving on the Editorial Boards for IEEE TRANSACTIONS ON GREEN COMMUNICATIONS AND NETWORKING, IEEE TRANSACTIONS ON COMPUTATIONAL SOCIAL SYSTEMS, and several other journals.
\end{IEEEbiography}

\begin{IEEEbiography}[{\includegraphics[width=1in,height=1.25in,clip,keepaspectratio]{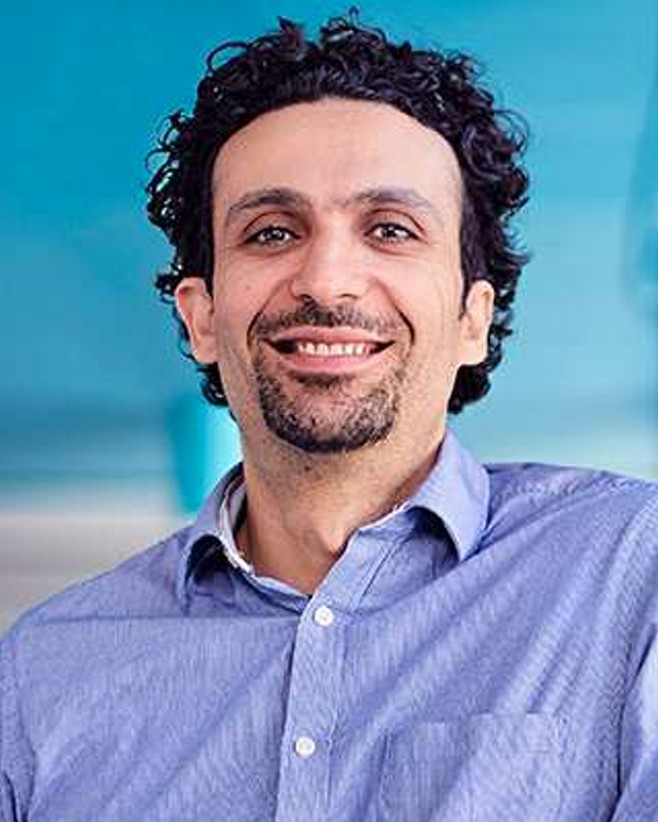}}]{Mehdi Bennis}
(Fellow, IEEE) is a Full (Tenured) Professor with the Centre for Wireless Communications, University of Oulu, Oulu, Finland, and the Head of the Intelligent Connectivity and Networks/Systems Group (ICON). His research interests include radio resource management, game theory and distributed AI in 5G/6G networks. He has published more than 300 research papers in international conferences, journals and book chapters. He has been a recipient of several prestigious awards including the 2015 Fred W. Ellersick Prize from the IEEE Communications Society, the 2016 Best Tutorial Prize from the IEEE Communications Society, the 2017 EURASIP Best paper Award for the Journal of Wireless Communications and Networks, the all-University of Oulu award for research, the 2019 IEEE ComSoc Radio Communications Committee Early Achievement Award, and the 2020–2026 Clarivate Highly Cited Researcher by the Web of Science.
\end{IEEEbiography}

\begin{IEEEbiography}[{\includegraphics[width=1in,height=1.25in,clip,keepaspectratio]{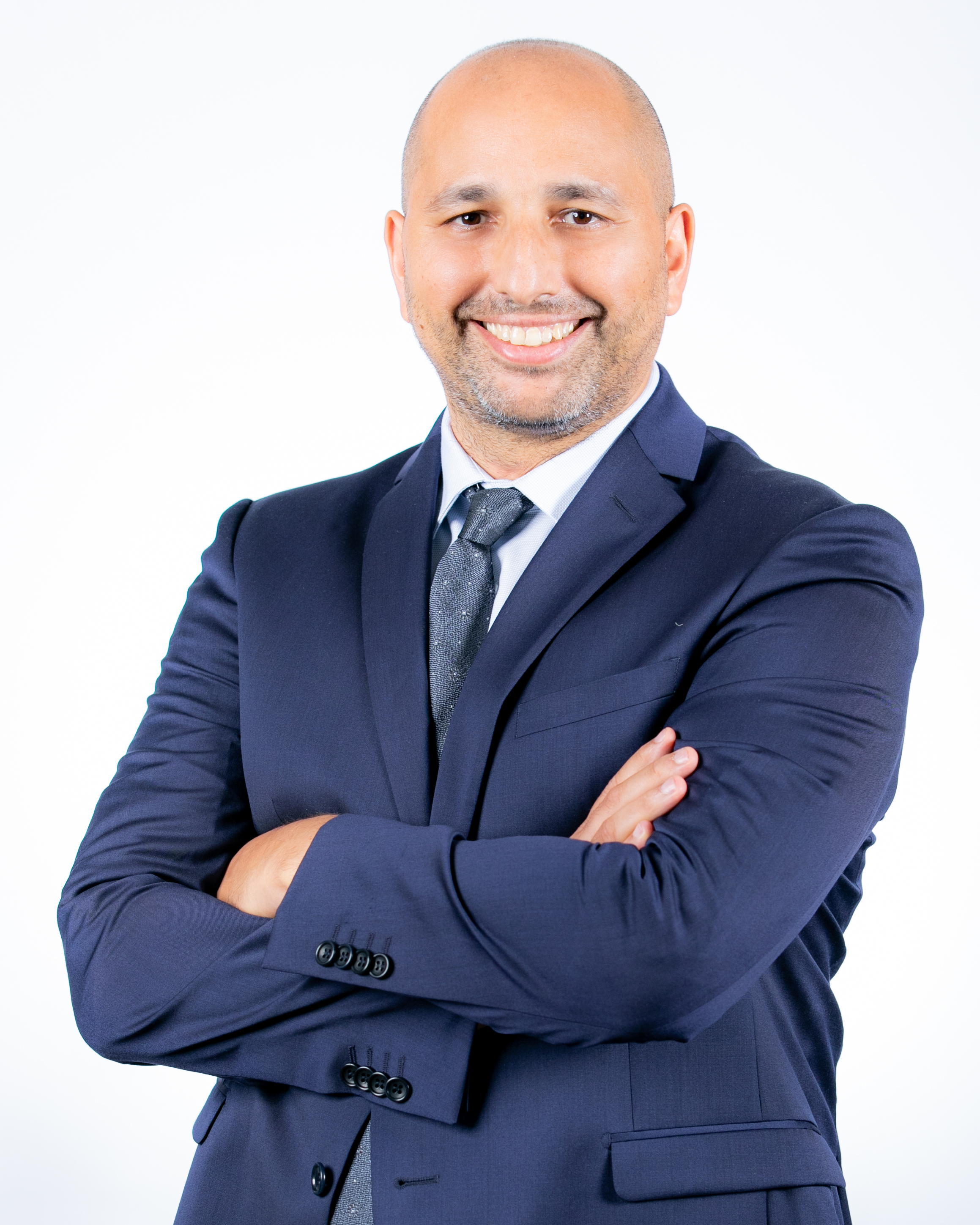}}]{M\'erouane Debbah}
(Fellow, IEEE) is Professor at Khalifa University of Science and Technology in Abu Dhabi and founding Senior Director of KU Digital Future Institute. His research has been lying at the interface of fundamental mathematics, algorithms, statistics, information and communication sciences with a special focus on random matrix theory and learning algorithms. In the Communication field, he has been at the heart of the development of small cells (4G), Massive MIMO (5G) and Large Intelligent Surfaces (6G) technologies. In the AI field, he is known for his work on Large Language Models, distributed AI systems for networks and semantic communications. He received multiple prestigious distinctions, prizes and best paper awards (more than 50 IEEE best paper awards) for his contributions to both fields. He is an IEEE Fellow, a WWRF Fellow, a Eurasip Fellow, an AAIA Fellow, an Institut Louis Bachelier Fellow, an AIIA Fellow and a Membre émérite SEE.
\end{IEEEbiography}

\begin{IEEEbiography}[{\includegraphics[width=1in,height=1.25in,clip,keepaspectratio]{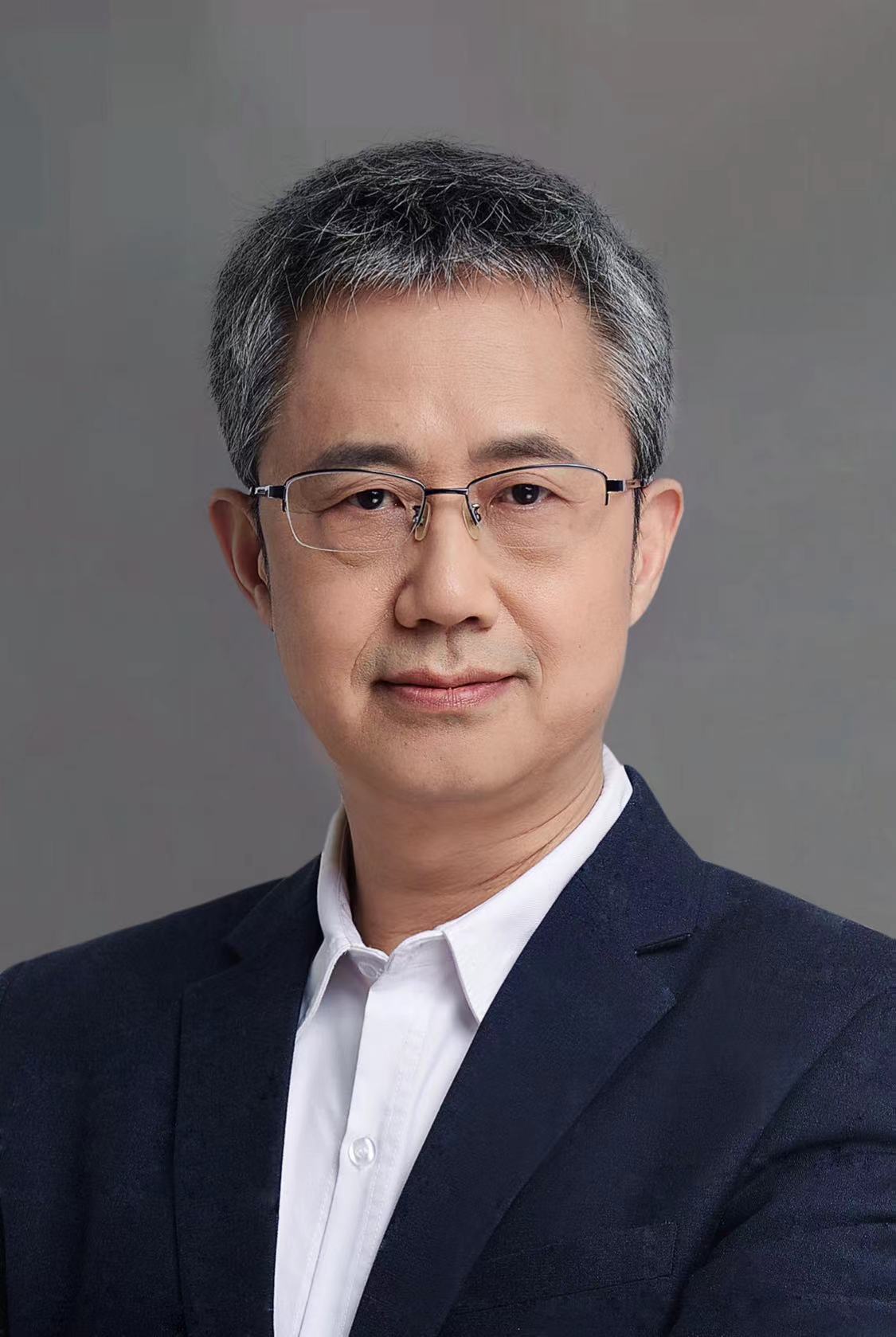}}]{Tie Jun Cui}
(M’98-SM’00-F’15) received the B.Sc., M.Sc., and Ph.D. degrees in electrical engineering from Xidian University, Xi’an, China, in 1987, 1990, and 1993, respectively. In March 1993, he joined the Department of Electromagnetic Engineering, Xidian University, and was promoted to an Associate Professor in November 1993. From 1995 to 1997 he was a Research Fellow with the Institut fur Hochstfrequenztechnik und Elektronik (IHE) at University of Karlsruhe, Germany. In July 1997, he joined the Center for Computational Electromagnetics, Department of Electrical and Computer Engineering, University of Illinois at Urbana-Champaign, first as a Postdoctoral Research Associate and then a Research Scientist. In September 2001, he became a Cheung-Kong Professor with the Department of Radio Engineering, Southeast University, Nanjing, China. Currently he is the Chief Professor of Southeast University, and the Director of State Key Laboratory of Millimeter Waves. He is also the Founding Director of the Institute of Electromagnetic Space, Southeast University. Dr. Cui’s research interests include metamaterials and computational electromagnetics. Dr. Cui is the Academician of Chinese Academy of Science, and IEEE Fellow. He served as Associate Editor of IEEE Transactions on Geoscience and Remote Sensing, and Guest Editors of Science China - Information Sciences, Science Bulletin, IEEE Transactions on Microwave Theory and Techniques, IEEE Journal of Emerging Technologies in Circuits and System, Applied Physics Letters, Engineering, Advanced Optical Materials, and Research. Currently he is the Chief Editor of Metamaterial Short Books in Cambridge University Press, the Senior Editor of IEEE Journal of Selected Topics in Electromagnetics, Antennas, and Propagation, the Editor of Materials Today Electronics, and the Editorial Board Members or International Advisory Members of National Science Review, eLight, PhotoniX, Small Structure, Physical Review Applied, npj Metamaterials, Research, Advanced Optical Materials, Advanced Photonics Research, and Journal of Physics: Photonics. He presented more than 200 Keynote and Plenary Talks in Academic Conferences, Symposiums, or Workshops.
\end{IEEEbiography}
\end{document}